\documentclass[]{aa}

\usepackage{amsmath}

\usepackage[colorlinks,breaklinks]{hyperref}
\hypersetup{linkcolor=blue,citecolor=blue,filecolor=black,urlcolor=blue}

\usepackage{amssymb}

\usepackage{xspace}
\usepackage{graphicx}
\usepackage{txfonts}
\usepackage{pdflscape}
\usepackage{comment}
\usepackage{orcidlink}
\usepackage{float}

\usepackage[switch]{lineno}

\begin{document} 

\title{ZTF SN Ia DR2: The diversity and relative rates of the thermonuclear supernova population}

\author{G.~Dimitriadis\inst{
\ref{trinity},\ref{lancaster}}\thanks{g.dimitriadis@lancaster.ac.uk}\orcidlink{0000-0001-9494-179X}
\and U.~Burgaz\inst{\ref{trinity}}\orcidlink{0000-0003-0126-3999}
\and M.~Deckers\inst{\ref{trinity}}\orcidlink{0000-0001-8857-9843}
\and K.~Maguire\inst{\ref{trinity}}\orcidlink{0000-0002-9770-3508}
\and J.~Johansson\inst{\ref{okcp}}\orcidlink{0000-0001-5975-290X}
\and M.~Smith\inst{\ref{ip2i},\ref{lancaster}}\orcidlink{0000-0002-3321-1432}
\and M.~Rigault\inst{\ref{ip2i}}\orcidlink{0000-0002-8121-2560}
\and C.~Frohmaier\inst{\ref{soton}}\orcidlink{0000-0001-9553-4723}
\and J. Sollerman\inst{\ref{okca}}\orcidlink{0000-0003-1546-6615}
\and L.~Galbany\inst{\ref{barca1},\ref{barca2}}\orcidlink{0000-0002-1296-6887}
\and Y.-L.~Kim\inst{\ref{lancaster}}\orcidlink{0000-0002-1031-0796}
\and C.~Liu\inst{\ref{northwestern},\ref{CIERA}}\orcidlink{0000-0002-7866-4531}
\and A.~A.~Miller\inst{\ref{northwestern},\ref{CIERA}}\orcidlink{0000-0001-9515-478X}
\and P.~E.~Nugent\inst{\ref{Lawrence},\ref{Berkeley}}\orcidlink{000-0002-3389-0586}
\and A.~Alburai\inst{\ref{barca1},\ref{barca2}}\orcidlink{0009-0007-2731-5562}
\and P.~Chen\inst{\ref{Weizmann}}\orcidlink{0000-0003-0853-6427}
\and S.~Dhawan\inst{\ref{Cambridge}}\orcidlink{0000-0002-2376-6979}
\and M.~Ginolin\inst{\ref{ip2i}}\orcidlink{0009-0004-5311-9301}
\and A.~Goobar\inst{\ref{okcp}}\orcidlink{0000-0002-4163-4996}
\and S.~L.~Groom\inst{\ref{IPAC}}\orcidlink{0000-0001-5668-3507}
\and L.~Harvey\inst{\ref{trinity}}\orcidlink{0000-0003-3393-9383}
\and W.~D.~Kenworthy\inst{\ref{okcp}}\orcidlink{0000-0002-5153-5983}
\and S.~R.~Kulkarni\inst{\ref{sri}}
\and K.~Phan\inst{\ref{barca1},\ref{barca2}}\orcidlink{0000-0001-6383-860X}
\and B.~Popovic\inst{\ref{ip2i}}\orcidlink{0000-0002-8012-6978}
\and R.~L.~Riddle\inst{\ref{caltech}}\orcidlink{0000-0002-0387-370X}
\and B.~Rusholme\inst{\ref{IPAC}}\orcidlink{0000-0001-7648-4142}
\and T.~E.~M{\"u}ller-Bravo\inst{\ref{barca1},\ref{barca2}}\orcidlink{0000-0003-3939-7167}
\and J. Nordin\inst{\ref{Berlin}}\orcidlink{0000-0001-8342-6274}
\and J.~H.~Terwel\inst{\ref{trinity},\ref{NOT}}\orcidlink{0000-0001-9834-3439}
\and A.~Townsend\inst{\ref{Berlin}}\orcidlink{0000-0001-6343-3362}
}

\institute{School of Physics, Trinity College Dublin, The University of Dublin, Dublin 2, Ireland \label{trinity}
\and Department of Physics, Lancaster University, Lancaster LA1 4YB, UK \label{lancaster}
\and Oskar Klein Centre, Department of Physics, Stockholm University, SE-10691 Stockholm, Sweden \label{okcp}
\and Universite Claude Bernard Lyon 1, CNRS, IP2I Lyon / IN2P3, IMR 5822, F-69622 Villeurbanne, France \label{ip2i}
\and Department of Physics and Astronomy, University of Southampton, Highfield, Southampton, SO17 1BJ, UK \label{soton}
\and Oskar Klein Centre, Department of Astronomy, Stockholm University, SE-10691 Stockholm, Sweden \label{okca}
\and Institute of Space Sciences (ICE, CSIC), Campus UAB, Carrer de Can Magrans, s/n, E-08193, Barcelona, Spain \label{barca1}
\and Institut d'Estudis Espacials de Catalunya (IEEC), E-08034 Barcelona, Spain \label{barca2}
\and Department of Physics and Astronomy, Northwestern University, 2145 Sheridan Rd, Evanston, IL 60208, USA \label{northwestern}
\and Center for Interdisciplinary Exploration and Research in Astrophysics (CIERA), Northwestern University, 1800 Sherman Ave, Evanston, IL 60201, USA \label{CIERA}
\and Lawrence Berkeley National Laboratory, 1 Cyclotron Road, MS 50B-4206, Berkeley, CA 94720, USA \label{Lawrence}
\and Department of Astronomy, University of California, Berkeley, 501 Campbell Hall, Berkeley, CA 94720, USA \label{Berkeley}
\and Department of Particle Physics and Astrophysics, Weizmann Institute of Science, 234 Herzl St, 7610001 Rehovot, Israel \label{Weizmann}
\and Institute of Astronomy and Kavli Institute for Cosmology, University of Cambridge, Madingley Road, Cambridge CB3 0HA, UK \label{Cambridge}
\and IPAC, California Institute of Technology, 1200 E. California Blvd, Pasadena, CA 91125, USA \label{IPAC}
\and Division of Physics, Mathematics and Astronomy, California Institute of Technology, Pasadena, CA 91125, USA \label{sri}
\and Department of Astronomy, California Institute of Technology, 1200 E. California Blvd, Pasadena, CA, 91125, USA \label{caltech}
\and Institut f{\"u}r Physik, Humboldt-Universit{\"a}t zu Berlin, Newtonstr. 15, 12489 Berlin, Germany \label{Berlin}
\and Nordic Optical Telescope, Rambla Jos{\'e} Ana Fern{\'a}ndez P{\'e}rez 7, ES-38711 Bre{\~n}a Baja, Spain \label{NOT}
}

\titlerunning{ZTF SN Ia DR2 - Photometric Diversity of SNe Ia}
\authorrunning{G. Dimitriadis et al.}

\abstract
{The Zwicky Transient Facility SN Ia Data Release 2 (ZTF SN Ia DR2) contains more than 3,000 Type Ia supernovae (SNe Ia), providing the largest homogeneous low-redshift sample of SNe Ia. Having at least one spectrum per event, this data collection is ideal for large-scale statistical studies of the photometric, spectroscopic and host-galaxy properties of SNe Ia, particularly of the rarer `peculiar' sub-classes. In this paper we first present the method we developed to spectroscopically classify the SNe in the sample, and the techniques we used to model their multi-band light curves and explore their photometric properties. We then show a method to distinguish between the peculiar sub-types and the normal SNe Ia. We also explore the properties of their host galaxies and estimate their relative rates, focusing on the peculiar sub-types and their connection to the cosmologically useful SNe Ia. Finally, we discuss the implications of our study with respect to the progenitor systems of the peculiar SN Ia events.}


\maketitle

%

\section{Introduction}
\label{sec:intro}

Modern observational cosmology relies heavily on the use of Type Ia SNe (SNe Ia) as a distance indicator for extragalactic distance measurements. Starting from the early 1990s, \citet{Phillips1993ApJ} used a sample of $\sim$10 SNe Ia to verify earlier suggestions \citep{Pskovskii1977SvA} of the standardisable nature of their light curves: the $B$-band absolute peak magnitude shows a strong correlation with their magnitude decline rate within 15 days after maximum ($\Delta m_{15}$), dubbed the width–luminosity relation (WLR). An increased sample of $\sim$30 SNe Ia additionally revealed a correlation between their peak magnitude and colour \citep{Riess1996ApJ}, establishing them as our most mature cosmological probes. The culmination of these studies led to the discovery of the accelerating expansion of the Universe due to `dark energy' \citep{Riess1998AJ,Perlmutter1999ApJ} using $\sim$20 and 40 SNe Ia at low ($z<0.1$) and high ($z>0.1$) redshifts respectively. This remarkable result has been corroborated with larger samples obtained during the 2000s \citep[e.g. 740 SNe Ia in][]{Betoule2014AA}.

As the current effort of SN Ia cosmological studies is focused on the nature of dark energy and precision measurements of its parameters, it was quickly realised that a shortcoming of previous attempts was the non-homogeneous set of low-redshift SN Ia observations. This need was addressed in the 2010s with the introduction of untargeted, high-cadence, low-redshift surveys, such as the Palomar Transient Factory \citep[PTF;][]{Law2009PASP,Rau2009PASP}, the All-Sky Automated Survey for Supernovae \citep[ASAS-SN;][]{Shappee2014ApJ}, the Panoramic Survey Telescope and Rapid Response System \citep[Pan-STARRS;][]{Chambers2016}, and the Asteroid Terrestrial-impact Last Alert System \citep[ATLAS;][]{Tonry2018PASP}. One of the main aims of these projects is the collection of high-quality SN Ia data at low redshift, in order to improve the standardisation techniques and reduce the systematic uncertainties that currently dominate the error budget \citep{Brout2022ApJ}. However, a surprising by-product was the realisation that SNe Ia do not constitute a uniform class of objects, as many `peculiar' sub-types that are photometrically outliers and/or show distinct spectroscopic properties have been found. While rarer, these events, if not identified and properly characterised, can contaminate the cosmologically useful SN Ia samples. Moreover, these peculiar transients are interesting on their own merits, as they challenge our understanding of the canonical SN Ia explosion paradigm. 

Theoretical models indicate that SN Ia progenitor systems contain a degenerate carbon-oxygen (C/O) white dwarf (WD) star undergoing explosive thermonuclear burning \citep{Hoyle1960ApJ} and a binary companion \citep{Whelan1973ApJ,Iben1984ApJS}. However, the nature of the companion is very poorly constrained, and can range from a non-degenerate companion to another degenerate WD, with many possible binary configurations and explosion mechanisms \citep{Liu2023RAA}. While (most) SN Ia progenitor scenarios generally reproduce the basic properties of the bulk of the thermonuclear SN population and, to some extent, their observational diversity, the question remains whether SNe Ia originate from a single progenitor scenario or if many paths towards a WD explosion do exist \citep{Jha2019NatAs}.

The observed photometric and spectroscopic diversity in the normal (i.e. the cosmologically useful) SNe Ia has been thoroughly investigated and is traditionally explained by the amount of radioactive $^{56}$Ni synthesised in the explosion \citep{Colgate1969ApJ}: an explosion of a degenerate non-rotating C/O WD close to the Chandrasekhar mass limit $\mathrm{M_{Ch}}$ \citep{Chandrasekhar31}, with larger $^{56}$Ni yields leading to hotter explosions with brighter and slower-declining light curves \citep{Kasen2007ApJ}. An alternative explanation involves explosions of WDs with masses lower than $\mathrm{M_{Ch}}$, where the total amount of ejecta mass sets the $^{56}$Ni yield and affects the SN luminosity and evolution timescale \citep{Goldstein2018ApJ,Shen2021ApJ}. The situation is more complicated for the peculiar events, as it is not yet clear if they represent extreme extensions of established progenitor scenarios or whether more exotic progenitor systems must be invoked \citep[see][for a recent review]{Taubenberger2017hsn}.

As different progenitor channels originate from different underlying stellar populations with different ages, estimating the SN production rate and understanding its dependence on the age distribution of its host stellar population provides a promising method to associate the different SN Ia sub-types with proposed progenitor scenarios. While considerable effort has been made for the normal events \citep{Frohmaier2019MNRAS,Perley2020ApJ}, estimating the rates of the peculiar events is challenging due to low number statistics, requiring a homogeneous survey with sufficient spectroscopic completeness \citep[see][for previous attempts]{Li2011MNRAS,Graur2017ApJ,Desai2024MNRAS}.

The aim of this paper is to collect and consistently study the peculiar thermonuclear SNe of the second data release of ZTF \citep[ZTF SN Ia DR2;][]{Rigault2024aAA}, investigate the properties of their host galaxies, and to estimate their production rates. In Sect.~\ref{sec:sample} we briefly describe the sample's photometry and spectroscopy, along with the additional metadata used in our analysis. In Sect.~\ref{sec:classifications} we outline the classification method we employed in order to classify the SNe in the sample. Section~\ref{sec:model_lc} discusses the modelling techniques we use in order to capture the photometric evolution of the SNe. In Sect.~\ref{sec:results} we present the results of our analysis, and we discuss these results in Sect.~\ref{sec:discussion}. Finally, we conclude in Sect.~\ref{sec:conclusion}. Throughout the paper, we adopt the AB magnitude system and a $\Lambda$CDM cosmology with a Hubble constant of $H_{0} = 73$ km s$^{-1}$ Mpc$^{-1}$.

\section{The ZTF SN Ia DR2 sample}
\label{sec:sample}

For this study we used data from the second data release of the Zwicky Transient Facility \citep[ZTF;][]{Bellm2019PASP,Graham2019PASP,Masci2019PASP,Dekany2020PASP}, acquired in the first three years of the survey \citep[2018 -- 2020;][]{Rigault2024aAA}. The result is an exceptional dataset of 3,628 spectroscopically classified events of the northern sky (Dec. $>-30\degr$). This sample corresponds to all ZTF events that have been flagged as SNe Ia in internal databases, such as GROWTH \citep{Kasliwal2019pasp} and Fritz \citep{vanderWalt2019, Coughlin2020apjs}, or public (mainly the Transient Name Server, TNS), for which at least one spectrum is of sufficient quality to classify the transient as a SN Ia (of any sub-type). The primary data include photometry in three optical bands ($ztf_{g}$, $ztf_{r}$ and $ztf_{i}$; hereafter we refer to them as $g$, $r$, and $i$ band, unless noted otherwise) and spectra obtained within the ZTF collaboration, with additional spectra from public sources (TNS) complementing the sample. Secondary data were collected over the course of the ZTF SN Ia DR2 compilation and include redshift estimates, sub-classification, light curve parameter fits, host galaxy parameters, and spectral indicators. In the following sections, we provide a brief summary of the dataset, while a detailed overview of the data acquisition and processing will be presented in \citet{smith2024AA}.

\subsection{Photometry}
\label{sec:phot}

Imaging observations of ZTF consist typically of 30s $g$, $r$, and $i$-band exposures, resulting in $5\sigma$ limiting magnitudes of $\sim20.6,\:20.4$, and $20.0$ in each filter. The typical cadence was 2--3 days in the $g$ and $r$ bands and 5 days in the $i$ band. We obtained force photometry light curves using the \texttt{ztflc} package with the \texttt{fpbot} interface \citep{reusch2023}. As the reference images in the ZTF pipeline are compiled from individual exposures taken during the survey, resulting in possible contamination of the transient's light in the reference image, we additionally employ a baseline correction, as discussed in \citet{smith2024AA}. Based on 
\citet{Rigault2024bAA}, we add a 2.5\%, 3.5\%, and 6\% error-floor in the $g$-, $r$-, and $i$-band fluxes respectively. The resulting light curve of each transient includes photometric points from March 2018 up to December 2020 (i.e. including non-detections prior to the first detection and after the transient has faded below the limiting magnitudes). 

\subsection{Spectra}
\label{sec:spec}

The ZTF SN Ia DR2 sample includes 5,138 optical spectra, acquired typically within 10 days to maximum brightness. The bulk of the spectra was obtained as part of the Bright Transient Survey \citep[BTS;][]{Fremling2020ApJ,Perley2020ApJ}, a magnitude-limited ($m<18.5$ at peak brightness) sample of extra-galactic transients in the ZTF public stream, with most of them ($\sim60\%$) acquired as classifications with the low-resolution SEDmachine \citep[SEDm;][]{Blagorodnova2018PASP,Rigault2019AA,Kim2022PASP}, mounted on the P60 telescope at Mount Palomar Observatory. The remainder of the spectra come from higher resolution instruments, such as the P200 Hale telescope, the New Technology Telescope (NTT), Keck and others, either as part of the BTS effort or from other resources. An overview of the spectra can be found in \citet{Johansson2024AA}, with additional information on the spectral reductions presented in \citet{Harvey2024AA}.

All spectra are fitted using the python wrapper \texttt{pysnid}\footnote{\url{https://github.com/MickaelRigault/pysnid}} of the Supernova Identification code \citep[\texttt{SNID};][]{Blondin2007ApJ}, employing a custom template library made of 370 templates (available upon request) that extends the variety of transient sources beyond the default templates. To perform the fit, we provide the known spectrum phase (given the ZTF light curves), and set the phase range to be the quadratic sum of the error on the maximum light $\delta t_{0}$ and a flat 4 days error to account for potential inaccuracy in the \texttt{SNID} template phase estimate.  We fit the 3,800$\,\AA$ to 10,000$\,\AA$ observer-frame wavelength range while discarding the first and last 10 wavelength pixels to avoid usual edge effects. We finally bound the redshift to be contained within $z\in[0,0.3]$ as ZTF cannot detect SNe~Ia at higher distances \citep[but see the lensed SNe~Ia SN Zwicky;][]{Goobar2023NatAs}. For some cases, a manual tuning of the wavelength range has been performed to help the fit to converge. At the end of this processing, only $0.8\%$ of our spectra did not have a \texttt{SNID} fit and were discarded from our analysis.

\subsection{Additional data}
\label{sec:metadata}

For each transient in the ZTF SN Ia DR2 sample we estimate its redshift, based on four methods. With increasing uncertainty, these are: a spectroscopic redshift obtained from public galaxy redshift catalogs, after correct determination of the host galaxy $(\delta z \leq 10^{-4})$, a spectroscopic redshift determined from visible host-galaxy emission lines in non-SEDm $(\delta z \leq 10^{-4})$ and SEDm spectra $(\delta z \leq 10^{-3})$, and an estimate of the redshift based on SNe Ia template matching ($\delta z \leq 3\times10^{-3}$). Most of our redshifts ($61\%$) are obtained from galaxy catalogs \citep[among which $71\%$ come from DESI through the Mosthost program;][]{Soumagnac2024ApJS}, $9\%$ from host-galaxy emission lines and $30\%$ from template matching. For the latter, we use the \texttt{SNID} fits from Sect.~\ref{sec:spec} and the transient's redshift is estimated considering only the top-matching templates of its sub-type (see Sect.~\ref{sec:classifications}). More information on the redshift estimates is included in \citet{Rigault2024aAA}.

All SN light curves have been fitted with the \texttt{SALT2} \citep{Guy2007AA} light-curve fitter, \citep[retrained model “SALT2-T21”;][]{Taylor2021MNRAS}, as implemented in sncosmo\footnote{\url{https://github.com/sncosmo/sncosmo}}, providing a measurement of the light curve stretch $x_{1}$ and colour parameter $c$, alongside the time of $B$-band maximum $t_{0}$. The \texttt{SALT2} fit parameters are available in \citet{Rigault2024aAA}.

The method with which the host galaxies of the ZTF SN Ia DR2 sample are identified and their parameters calculated is thoroughly presented in \citet{smith2024AA}. In short, we firstly determine all the galaxy sources within a fixed radius of 200 kpc (determined from the estimated redshift of the SN) centred on each SN location from querying (in priority order) the DESI Legacy Imaging Survey \citep[DESI-LS;][]{Dey2019AJ} DR9, the Sloan Digital Sky Survey (SDSS) DR17 \citep{Abdurrouf2022ApJS}, and the Pan-STARRS \citep[PS1;][]{Chambers2016} DR2 catalogs. Then, the software package \texttt{HostPhot}\footnote{\url{https://github.com/temuller/hostphot/tree/main}} \citep{Muller-Bravo2022} was applied in the PS1 images, in order to calculate, for each potential host galaxy, their elliptical parameters, Kron fluxes and Directional Light Radii \citep[$d_{DLR}$;][]{Sullivan2006APJ, Smith2012ApJ,Sako2018PASP}, the (normalised by the elliptical radius of the galaxy in the direction of the SN) separation of the SN with the host. Finally, the galaxy source (within a $d_{DLR}<7$ range) with the minimum $d_{DLR}$ was selected as the matched host. Final photometry for each host galaxy, and, additionally, to the location of the SN in a 2 kpc radius, was performed in PS1 $g,r,i,z$, and $y$ bands. In ZTF SN Ia DR2, we provide rest-frame $g-z$ colour and total stellar mass estimates by fitting the spectral energy distribution with a custom fitting code also applied in \citet{Sullivan2010MNRAS} and using the \texttt{P\'{E}GASE.2} templates \citep{Fioc1997AA}.

For the majority of the spectra, spectroscopic indicators, such as the line expansion velocities and pseudo-equivalent widths of the \ion{Si}{ii} $\lambda6355$ and \ion{Si}{ii} $\lambda5972$, are derived with the method of \citet{Childress2013ApJ}, in which multiple Gaussians (to account for the doublet nature of the silicon lines) are fitted (in velocity space) in the continuum-normalised region of the relevant line. A detailed presentation of this method and its application in a sub-sample of the ZTF SN Ia DR2 is presented in \citet{Burgaz2024AA}.

\section{Classifications of the ZTF SN Ia DR2 sample}
\label{sec:classifications}

In order to efficiently (sub-)type 3,628 SNe Ia with 5,138 spectra, a combination of an automated classification software with the input of multiple SNe Ia expert classifiers was adopted. This process was made in two steps: first, general user input made through a specially designed web-application; second, careful fine-tuned updates providing spectra and light curve features.

\subsection{\texttt{“Typingapp”} as classification input}
\label{sec:typingapp}

\begin{figure*}
    \centering
    \includegraphics[width=2\columnwidth]{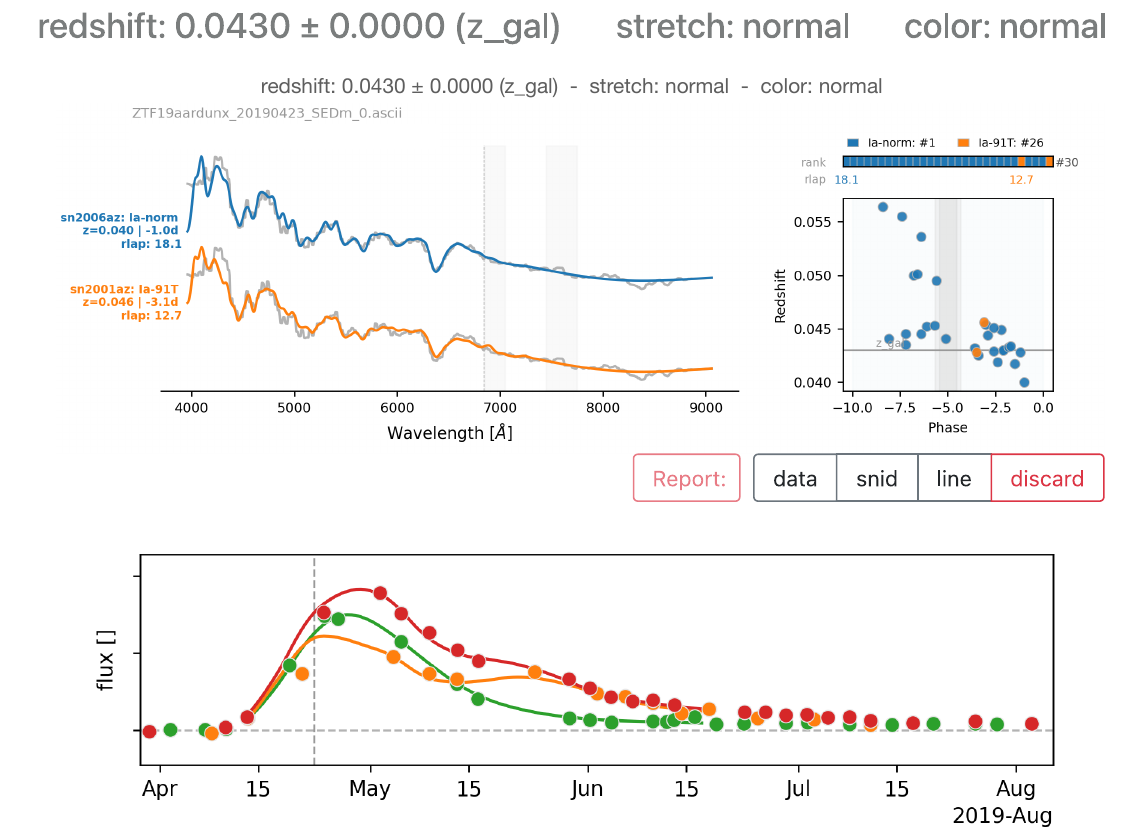}
    \caption{Example of a \texttt{typingapp} webpage for ZTF19aardunx (SN\,2019dpw), a typical SN Ia at $z=0.043$. Its spectrum is fitted with \texttt{SNID} and (minimal) light curve fit information is displayed at the top of the figure. \textbf{Top left}: Target spectrum (in grey; the two curves are the same) overplotted with the best matching (rlap) entry (sub-)type template spectra, considering only the best 30 \texttt{SNID} entries (see top right panel). The vertical grey bands show the main telluric line locations. \textbf{Top right}: Summary of the \texttt{SNID} 30 best entries sorted by rlap. The lower panel shows the template matching redshift-phase correlations, while the top panel shows the entry (sub-)type sorting. The best rlap per (sub-)entry corresponding to the spectrum displayed on the left is shown. The grey vertical bands show the spectrum's actual phase ($\pm 2\sigma$) given the lightcurve's $t_{0}$ estimation. This target had five user classification inputs, three as snia-norm and two as snia, and thus it was automatically classified as a snia-norm. \textbf{Bottom:} Target light curve in $g$, $r$, and $i$ band, shown as green, red, and orange circles, respectively, alongside the best fitted \texttt{SALT2} model. The vertical dashed line corresponds to the date of the spectrum. The values and units of the transient's flux are not displayed in order to avoid biasing the classifier.}
    \label{fig:typping_example}
\end{figure*}

We developed the \texttt{typingapp} web-application\footnote{\url{https://typingapp.in2p3.fr/dashboard}} where users were invited to (sub-)classify the data given their spectra with the \texttt{SNID} fit results (see Sect.~\ref{sec:spec}), as illustrated in Fig.~\ref{fig:typping_example}. In addition, minimal information was given to the user, such as the redshift (and its origin), and minimal light curve parameter information such as if the \texttt{SALT2} stretch and colour had been flagged as `normal' or not. No host information nor detailed parameters were displayed in this front-end to avoid biasing the users.

Given the information displayed, users were invited to classify targets as `snia', `non-snia' or `unclear'. If possible, snia could further be sub-classified as `snia-norm', `snia-91T', `snia-91bg' or `snia-other'. Along the process, users were also invited to flag bad spectra, badly fitted light curves or spectra, or point out existing potential host-emission lines that could be used to estimate host redshifts (see Sect.~\ref{sec:metadata}). A link to the internal GROWTH and Fritz database front-end was also provided, if users needed additional information.

For this data release, more than 14,000 (sub-)classifications were made by 32 different users. On average, each target has been (sub-)classified 3.5 times and all at least twice. At the end of the process, all individual classifications have been gathered and the final (sub-)typing could have three different sources: `auto': at least three users agree on the target (sub-)classification; `expert': at least two experts agree on the target (sub-)classification; `arbiter', this target was manually double checked and vetted by experts. The following enumeration explains in detail how this procedure was made. If a target had several entries, arbiter supersedes expert that supersedes auto. 

Given the \texttt{typingapp} classifications, the auto classification is performed as follows:
\begin{enumerate}
    \item If all (sub-)classifications agree, it defines its typing. About 30\% of the sample is classified this way; two-thirds as snia-norm, one-quarter as snia.
    \item Else, if classifications are combinations of snia-norm and snia, it is considered as snia-norm.
    \item If all classifications agree on the Ia Type but disagree with the sub-classification (i.e. snia-norm, snia-91T, snia, snia-norm], the classification is send to the arbiter.
    \item Else, if all classifications agree on the non-snia type, the target is not considered as a SN Ia.
    \item Else, if the unclear classification accounts for fewer than half of the entries, the procedure restarts removing the unclear entries.
    \item Else, the classification is send to the arbiter.
    \item The expert classification follows the exact same procedure but only accounts for entries from a selected list of expert users; then only two classification inputs are enough.
\end{enumerate}

At the end of this process, 34 events ($\sim1$\% of the initial sample) were determined to not belong on the SN Ia class and were excluded from further analysis. For the events identified as SN Ia, 32\% of the sample had auto and 44\% expert classifications. The 24\% remaining confusing cases that had to be inspected by the arbiters were mostly non-normal SNe~Ia.

\subsection{Final (sub-)classification}
\label{sec:subclassification}

The results of the \texttt{typingapp} were used as input to a more thorough sub-classification. Firstly, the events typed as snia-other were further characterised, according to their peculiar sub-type. This list contains 78 SNe and their sub-classifications were based on their \texttt{SNID} fit results and (for a sub-set of them) studies already performed and published \cite[e.g.][]{De2019ApJ,Tomasella2020MNRAS,Miller2020ApJ,Srivastav2020ApJ,Hoeflich2021ApJ,Karambelkar2021ApJ,Dimitriadis2022ApJ,Dutta2022ApJ,Singh2022MNRAS,Srivastav2022MNRAS,Dimitriadis2023MNRAS,Kool2023Natur,Maguire2023MNRAS,Sharma2023ApJ,Liu2023ApJ}. The peculiar sub-types considered were the following:

\begin{itemize}
    \item The overluminous and slow evolving `03fg-like' SNe Ia \citep{Howell2006Natur}. This class of objects is characterised by their increased peak brightness relative to their (broad) light curves, their low ejecta velocity and the presence of strong and persistent \ion{C}{ii} $\lambda6580$ and $\lambda7234$. We note that, throughout the literature, various monikers for this class have been used, such as super-Chandrasekhar-mass SNe Ia, 09dc-like, (carbon-rich) overluminous SNe Ia and superluminous SNe Ia \citep{Taubenberger2011MNRAS,Ashall2021ApJ}.
    \item The underluminous and spectroscopically unusual SNe Iax. These transients have lower luminosities and ejecta velocities, with peculiar late-time spectra, as they never become fully nebular. Initially dubbed as `02cx-like', based on the prototypical SN\,2002cx \citep{Li2003PASP}, SNe Iax are thought to be the most common peculiar sub-class of thermonuclear SNe \citep{Foley2013ApJ}.
    \item The interacting SNe Ia-CSM. These events show evidence of strong interaction of their ejecta with hydrogen-rich circumstellar medium (CSM). While there are cases where a strong interaction conceals the underlying SN Ia signatures \citep[e.g.][for SN\,2002ic]{Benetti2006ApJ}, suggesting a core-collapse explosion, events with weaker interaction \citep[e.g.][for PTF~11kx]{Dilday2012Sci} strengthen their thermonuclear origin \citep{Silverman2013ApJS}.
    \item The underluminous but with normal light-curve timescales `02es-like' SNe Ia \citep{Ganeshalingam2012ApJ}. These events share many characteristics with the sub-luminous SNe Ia, featuring cool ejecta with low ionization spectra; however, their light curve evolution is somewhat slower. We note that, as this is a recently defined sub-class, the actual member definition is not well established, but rather refers to objects that occupy a region in the WLR space previously thought to be empty.
    \item Finally, we include a new sub-class, the `18byg-like' SNe Ia \citep{De2019ApJ}. These events resemble the sub-luminous SNe Ia in terms of their peak brightness, have strong red colors due to a significant flux suppression in their bluer part of the spectrum before the SN reaches maximum luminosity and show indications of helium burning in their spectra, such as \ion{Ti}{ii} and \ion{Ca}{ii} \citep{Inserra2015ApJ}.
\end{itemize}

After the \texttt{typingapp} results were collected, a concentrated sub-typing effort was performed for the SNe Ia in the volume-limited (for normal SNe~Ia) sample \citep[$z\leq0.06$;][]{Amenouche2024AA}, containing 1,583 SNe. In this sample, 193 events did not have sufficient spectral quality to determine their sub-type, thus remained as `not sub-typed', and 46 events were identified as peculiar. The remaining 1,344 SNe correspond to the normal SN~Ia population alongside the traditional peculiar sub-classes of the overluminous `91T-like' \citep{Phillips1992AJ} and underluminous `91bg-like' SNe Ia \citep{Filippenko1992AJ}. Their sub-type was determined by careful visual inspection, spectral comparisons with well-observed historical SNe and measurements of spectral indicators. In particular, we followed \citet{Phillips2022ApJ} and calculated the pEW of the \ion{Si}{ii} $\lambda$6355 line as a function of phase, in order to differentiate between normal and 91T/99aa-like events, and the \ion{Si}{ii} $\lambda$5972$/$$\lambda$6355 strength ratio and the line complex at the (rest-frame) 4,300 \AA\ wavelength region where strong \ion{Ti}{ii} is observed for the low-luminosity 91bg-like SNe Ia.

\begin{table}
\caption{Final (sub-)classifications of the ZTF SN Ia DR2 sample.}
\centering
\label{tab:class_numbers}
\begin{tabular}{l c}
\hline\\[-0.5em]
    Sub-class & No. of SNe \\
    \hline\\[-0.8em]
    \hline\\[-0.5em]
    Complete sample & \textbf{3628} \\
    \hline\\[-0.5em]
    Normal & 2511 \\
    91T-like & 292 \\
    91bg-like & 92 \\
    03fg-like & 29 \\
    Iax & 23 \\
    Ia-CSM & 14 \\
    02es-like & 8 \\
    18byg-like & 4 \\
    Not sub-typed & 655 \\
    \hline\\[-0.5em]
    Volume limited & \\
    ($z\leq0.06$) & \textbf{1583} \\
    \hline\\[-0.5em]
    Normal & 1113 \\
    91T-like & 145  \\
    91bg-like & 86 \\
    03fg-like & 11 \\
    Iax & 20 \\
    Ia-CSM & 5 \\
    02es-like & 7 \\
    18byg-like & 3 \\
    Not sub-typed & 193 \\
    \hline\\[-0.5em]
\end{tabular}
\tablefoot{See text in Sect.~\ref{sec:classifications} for a discussion of specific sub-types and individual events.}
\end{table}

\begin{figure*}
    \centering
    \includegraphics[width=2\columnwidth]{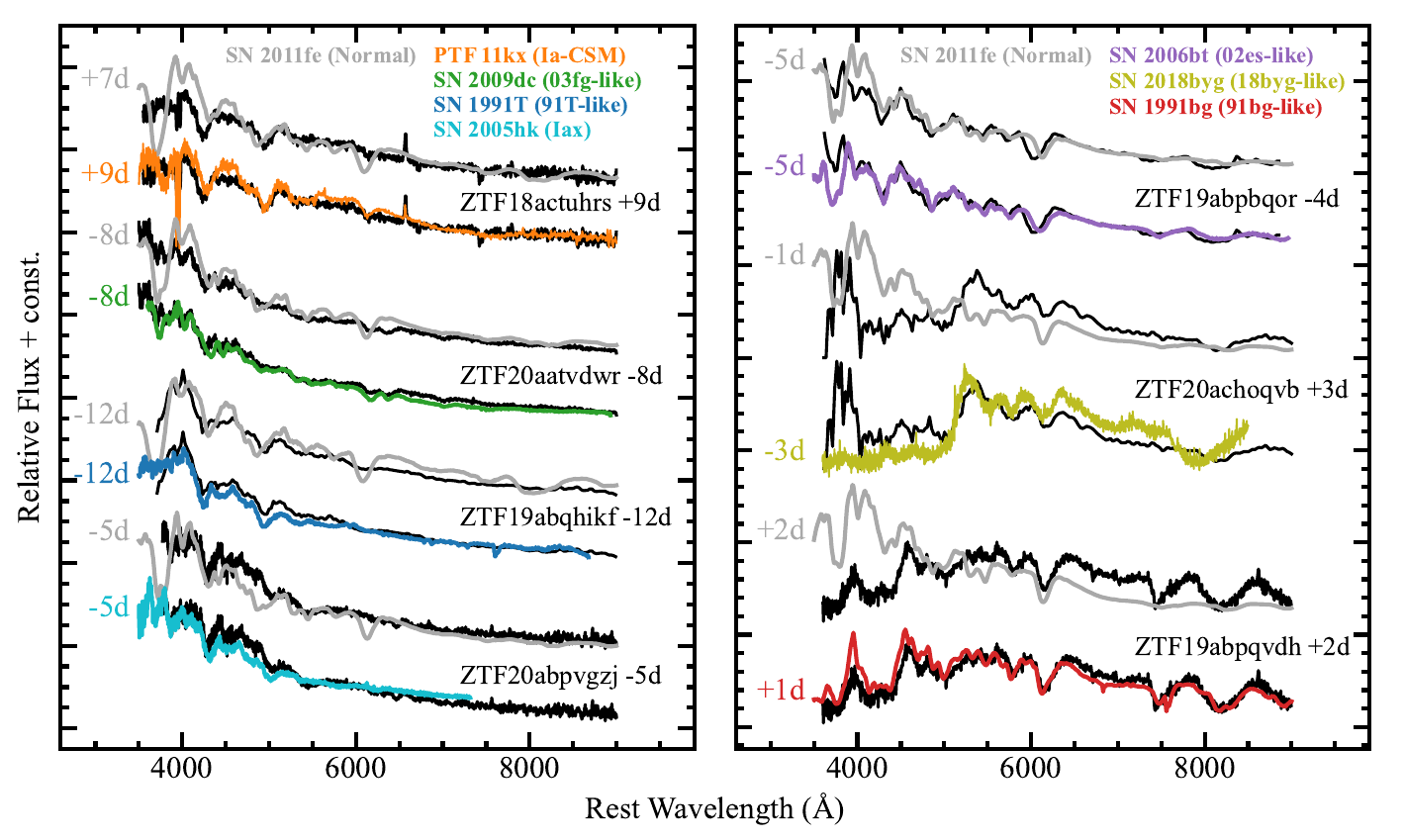}
    \caption{Collection of non-normal ZTF SN Ia DR2 and their spectra, compared with spectra of the normal SN Ia 2011fe from \citet{Pereira2013AA} and their matched peculiar sub-types: SNe Ia-CSM \citep[PTF\,11kx;][]{Dilday2012Sci}, 03fg-like \citep[SN\,2009dc;][]{Taubenberger2011MNRAS}, 91T-like \citep [SN\,1991T;][]{Mazzali1995AA}, SNe Iax \citep[SN\,2005hk;][]{Blondin2012AJ}, 02es-like \citep[SN\,2006bt;][]{Foley2010ApJ}, 18byg-like \citep[SN\,2018byg;][]{De2019ApJ} and 91bg-like \citep[SN\,1991bg;][]{Filippenko1992AJ}. We additionally mark the time from maximum (in rest-frame days) of each spectrum, with the exception of SNe Ia-CSM, where we mark the time from explosion.}
    \label{fig:peculiar_spectra}
\end{figure*}

The results of our classification effort are presented in Table~\ref{tab:class_numbers}. A collection of spectra of various peculiar events, demonstrating the spectroscopic diversity of the thermonuclear explosions, is shown in Fig.~\ref{fig:peculiar_spectra}. Below, we comment on certain aspects which affect our typing and discuss individual interesting events.

\begin{itemize}
    \item Our primary tool for (sub-)classifying a SN Ia is the templates included in \texttt{SNID}. While \texttt{SNID} and similar software, such as \texttt{SuperFit} \citep{Howell2005ApJ} and \texttt{GELATO} \citep{Harutyunyan2008AA}, are still widely used and perform decent SNe Ia classifications, their spectral templates generally do not take into account recently identified sub-classes (e.g. SN\,2002es is typed as 91bg-like in \texttt{SNID}). By employing a multiple-users approach, and with the involvement of SN~Ia population experts, we believe that our typing is sufficiently accurate.
    \item \citet{Burgaz2024AA} present a more detailed analysis of the diversity in the normal SNe Ia population and the two transition regions of normal to overluminous and normal to sub-luminous, additionally considering 99aa-like \citep{Garavini2004AJ} and 86G-like \citep{Phillips1987PASP} events, while introducing the `04gs-like' sub-class. In this study, we will include 99aa-like events in the 91T-like, 86G-like in the 91bg-like and 04gs-like in the normal sub-class, respectively.
    \item \citet{Burgaz2024AA} have discussed the impact of the host galaxy contamination on the spectra for the (sub-)classification of SNe Ia, and particularly on the strength of the \ion{Si}{ii} $\lambda$6355 absorption line, a defining characteristic of SNe Ia and a spectral indicator used to differentiate between normal and 91T/99aa-like events. In our sample, this results in a trend where most of the not sub-typed classifications correspond to SNe with small separation from the host galaxy's core, thus larger contribution of the galaxy light in the spectra.
    \item The majority of our spectra ($\sim$67\%) were taken with SEDm and SPRAT, posing difficulties resolving spectral features crucial in identifying peculiar sub-types (e.g. the \ion{C}{ii} $\lambda6580$ line in 03fg-like), due to the low-resolution or the limited wavelength coverage of these spectrographs. This potentially has the effect of mis-classifying peculiar events as normal, and reducing the total number of the peculiar population.
    \item For most of thermonuclear explosions, the spectroscopic differences between the sub-classes are observed at early times, while at $\sim$10--15 days from peak brightness their spectra appear relatively similar. This problem affects mainly the 91T-like (and their transitional sub-types), the 03fg-like and, to some extent, the SNe Iax sub-classes. Most of our spectra ($\sim$73\%) were taken $<$10 days from maximum, alleviating this issue. Alternatively, multiple spectra per object may provide a better insight, but the majority of our sample ($\sim$72\%) has one spectrum per event.
    \item For most of the non-normal sub-types of SNe Ia, an intrinsic diversity has been observed (see e.g. \citet{Taubenberger2019MNRAS} and \citet{Ashall2021ApJ} for 03fg-like, and \citet{Taubenberger2017hsn} for 02es-like). In this study, we refrain from investigating this diversity and consider homogeneous sub-classes.
    \item Three events, ZTF18acwvdoo (SN\,2018kax), ZTF19aamgypy (SN\,2019bka) and ZTF20aahidph (SN\,2020aqa) were identified as 00cx-like \citep{Li2001PASP}, an extremely rare sub-type of overluminous SNe Ia that shares many characteristics with the 91T-like SNe Ia. While there is an indication that these objects are distinct from the 91T-like (primarily due to the local environment in which they occur), we include them in the 91T-like sub-class.
    \item ZTF20aatxryt (SN\,2020eyj) is the first SN Ia-CSM with He-rich and H-poor circumstellar medium \citep{Kool2023Natur}. While possibly distinct, we include it in the broader SN Ia-CSM sub-class.
    \item ZTF20abrjmgi (SN\,2020qxp) is classified in \citet{Hoeflich2021ApJ} as a low-luminosity event, with a convincingly 91bg-like early ($\sim$ 10 days before maximum) spectrum. While we lack $g$-band coverage of the event, and our $r$- and $i$-band light curve is limited to $\sim$ 15 days after maximum, the SN has a slow evolution ($x_{1}=0.675$), relative to 91bg-like events. Thus, we tentatively classify it as an 02es-like SN Ia.
    \item ZTF19adcecwu (SN\,2019yvq) was identified as a peculiar thermonuclear explosion by \citet{Miller2020ApJ}, and has been extensively studied \citep{Siebert2020ApJ,Tucker2021ApJ}. While we do acknowledge its unique features, we include it in the 02es-like sub-class, as it was tentatively classified as such by \citet{Burke2021ApJ}.
\end{itemize}

\section{Modelling the light curves of the ZTF SN Ia DR2 sample}
\label{sec:model_lc}

In this section, we describe the sub-type specific K-corrections we apply to each SN (Sect.~\ref{sec:kcor}) and the Gaussian Processes (GPs) light curve fits we employ to model each event's photometric evolution (Sect.~\ref{sec:gpfits}). Finally, we describe the final samples we use in our analysis to investigate the thermonuclear SN diversity in Sect.~\ref{sec:final_samples}.

\subsection{K-corrections}
\label{sec:kcor}

In order to consistently compare the intrinsic photometric properties of a diverse sample of sources at varying distances, it is crucial to correct for the effects of redshift between observed and rest-frame broad-band photometry (i.e. the relation between the rest-frame absolute magnitude to the observed-frame apparent magnitude of a source as a function of time, in a broad photometric bandpass). This set of processes, known as K-corrections \citep{Kim1996PASP,Nugent2002PASP}, are routinely used in cosmological studies of SNe Ia, particularly when comparing nearby and distant SNe, where an observed-frame light curve in a specific bandpass corresponds to a rest-frame light curve in a bluer bandpass. However, in most studies that mainly focus on SNe Ia physics, K-corrections are omitted or roughly approximated, usually due to the low-redshift nature of the studied sample (for example the CSP-II \citep{Ashall2020ApJ} and the ASAS-SN \citep{Desai2024MNRAS} samples had a median redshift of $0.024$).

The most challenging problem when applying K-corrections is that usually there is no simultaneous photometric and spectroscopic observations of the transient, thus, spectroscopic templates are commonly used (e.g. \citealt{Hsiao2007ApJ,Guy2007AA}). This choice is further justified as, at least for the normal SNe Ia, the spectral time evolution is remarkably homogeneous, with small differences attributed to the intrinsic diversity in the light curve shape, for which most spectral templates account for. However, complications arise when considering the peculiar SN Ia population, as their spectral energy distribution (SED) evolution could differ, for example the lower line velocities of SNe~Iax \citep{Foley2013ApJ} or the presence of strong hydrogen emission lines in SNe Ia-CSM \citep{Silverman2013ApJS}. As some of these rare sub-types tend to be discovered at higher redshifts, the effect of K-corrections must be taken into account.

In this study, we attempt to K-correct all the SN Ia light curves in our sample, since the ZTF SN Ia DR2 sample's redshift distribution peaks at higher values compared to previous studies ($z_\mathrm{{med}}=0.07$). All light curves are K-corrected with the use of the \texttt{SALT2} spectroscopic template, which is based on each event's estimated light curve stretch $x_{1}$, colour parameter $c$, time of maximum (in the B-band) and redshift. This method is justifiable for the normal SNe Ia and, to some extent, the overluminous 91T-like SNe Ia, as \texttt{SALT2} provides decent light curve fit results, and the spectroscopic peculiarities are generally captured in the template. While this approach does not take into account the complete spectroscopic diversity of the rest of the sub-types, it nevertheless provides a consistent framework. We also use this method for the events for which we were unable to sub-classify (not sub-typed).

For the remainder of the peculiar sub-types, and in addition to the \texttt{SALT2} template, we use:

\begin{itemize}
    \item For the underluminous SNe Ia sub-type of 91bg-like, we use the spectroscopic template `SN Ia-91bg' from the Photometric LSST Astronomical Time Series Classification Challenge (PLAsTiCC), presented in \citet{Kessler2019PASP}. This template is a set of 35 SED template series, based on the original \citet{Nugent2002PASP} template, that covers a grid of \texttt{SiFTO} \citep{Conley2008ApJ} stretches and colours. We fit all our 91bg-like SNe Ia ZTF light curves with \texttt{SiFTO}, estimate their stretch, colour and time of maximum, and select the closest template at the stretch/colour space to apply the K-corrections.
    \item For the peculiar sub-type of 02es-like, we use the same method as with the 91bg-like, as these events are spectroscopically similar to 91bg-like, but with slower evolution, a characteristic that is encompassed in the \texttt{SiFTO} stretch. A notable exception is ZTF19adcecwu, for which we do not attempt to K-correct, as it occurred at a very low redshift ($z=0.009$).
    \item For the peculiar sub-type of SNe Iax, we use the spectroscopic template `SN Iax' from PLAsTiCC. This template is based on observations of the well-observed SN\,2005hk \citep{Chornock2006PASP,Phillips2007PASP,Sahu2008ApJ} and it is adapted to match the luminosity function and light curve parameters for a population of SN Iax. This template is a set of 1000 SED template series that cover a grid of absolute peak luminosity and rise time in V-band, and decline rates in the B- and R-bands (which are all correlated). We perform an initial fit to all our SNe Iax ZTF light curves and determine a preliminary absolute peak $g$-band magnitude. Since the $g$ band is sufficiently close to V-band, we collect all the template spectra that correspond to $\pm0.1$ of the estimated peak magnitude, we average them for each phase bin and use them to apply the K-corrections.
    \item For the peculiar sub-type of 03fg-like, and since no spectroscopic templates have been constructed yet, we use the (spectrophotometrically calibrated) spectral series of the well-observed SN\,2009dc from \citet{Taubenberger2011MNRAS}.
    \item For the peculiar sub-type of 18byg-like, we use our spectral series of the prototypical ZTF18aaqeasu (SN\,2018byg). We first mangle the spectra from \citet{De2019ApJ} to the ZTF photometry, calculate the K-corrections at the spectral epochs and then, as the spectral evolution appears to be smooth, we interpolate with GP fits and apply the K-corrections.
    \item For the peculiar sub-type of the SNe Ia-CSM, we chose not to apply any K-corrections to the data, as no spectroscopic templates have been constructed yet, and their spectroscopic evolution and intrinsic diversity appear to be too complex to be covered in a single template.
\end{itemize}

We note that, for the 03fg and 18byg-like sub-classes, using a single, albeit well-observed and fairly representative object for calculating the K-corrections prevents us from encompassing a possible intrinsic diversity in its class (e.g. the spectral evolution of 18byg-like events is probably sensitive to the amount of Fe-group elements synthesised during the helium-shell detonation).

\begin{figure}
    \centering
    \includegraphics[width=1\columnwidth]{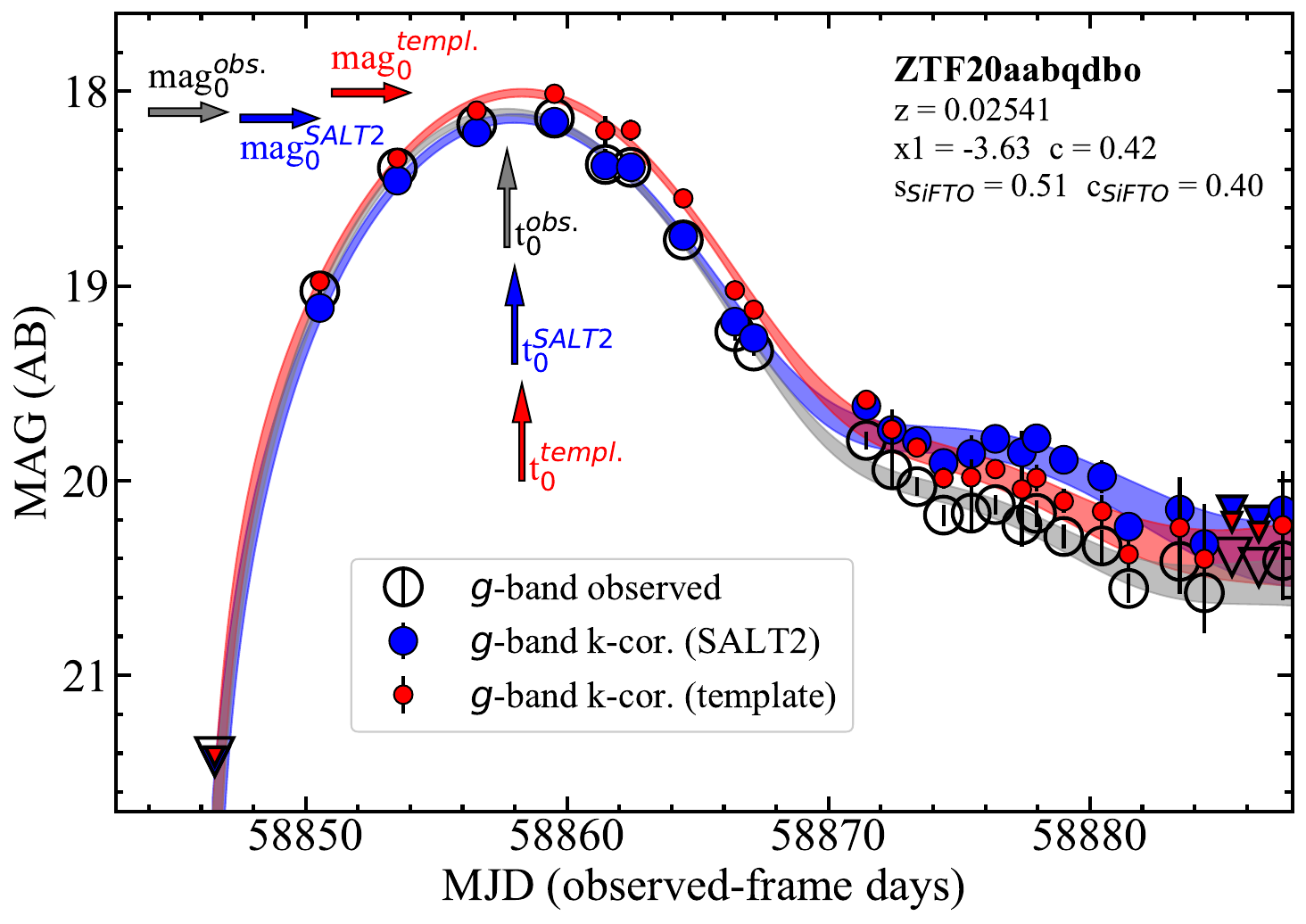}
    \caption{Observed $g$-band light curve of the 91bg-like ZTF20aabqdbo (SN\,2020dg, open black circles; the downward triangles refer to non-detections) in observed days, compared to the same light curve, but K-corrected based on the \texttt{SALT2} (blue circles) and on the PLAsTiCC template (red circles). The transient's redshift, \texttt{SALT2}'s stretch $x_{1}$ and colour parameter $c$, and \texttt{SiFTO}'s stretch $s$ and colour parameter $c$ are shown in the legend. The shaded regions correspond to GP fits of the light curves. We note the difference of the estimated time of (upward arrows) and magnitude (right-pointing arrows) at maximum for each light curve.}
    \label{fig:kcor_comparison}
\end{figure}

Fig.~\ref{fig:kcor_comparison} shows the effect of the application of different K-correction templates to the observed (in this case, the $g$ band) light curve for the case of a 91bg-like ZTF SN Ia. As can be seen, many of the light curve parameters we estimate (e.g. the time of maximum, the peak magnitude, and the decline rate for each photometric band) are directly affected. In particular, we find that, for the 91bg-like events, the effect is more prevalent in the $g$ band, for which the template-calculated peak brightness is $\sim0.15$ mag brighter and the +15 d decline rate is $\sim0.22$ mag faster than the salt-calculated ones. The effect is milder in the 03fg-like ($\sim0.03-0.1$ mag) and the SNe Iax ($\sim0.04-0.1$ mag), but still notable, given that the uncertainty in these parameters are in the order of $\sim0.05-0.1$ mag.

\subsection{Gaussian Processes light curve fits}
\label{sec:gpfits}

Our main tool for modelling a SN light curve in order to characterise its photometric evolution and investigate the diversity in our sample is the GP Regression \citep[see][for a comprehensive discussion of GPs and their applications]{Rasmussen2006gpml}. GPs light curve fits have been widely used, particularly for photometric classification purposes \citep[e.g.][]{Lochner2016ApJS,Revsbech2018MNRAS,Boone2019AJ} and for determining cosmological distances \citep[e.g.][]{Kim2013ApJ,Saunders2018ApJ,Muller2022MNRAS}. Our motivation to use GPs instead of other methods of (template-based) light-curve fitting is based on our focus on the peculiar events. Most of these codes, such as \texttt{SALT2}, \texttt{SiFTO} \citep{Conley2008ApJ} or \texttt{SNooPy} \citep{Burns2011AJ} perform optimally only with events that resemble their training sample, typically well-observed, low-redshift and normal SNe~Ia, naturally failing to accurately determine the properties of non-normal events. With our GPs approach, no underlying template or model is used in the fit, thus the analysis of each transient's multi-wavelength light curves is performed in a purely data-driven basis.

\begin{figure}
    \centering
    \includegraphics[width=1\columnwidth]{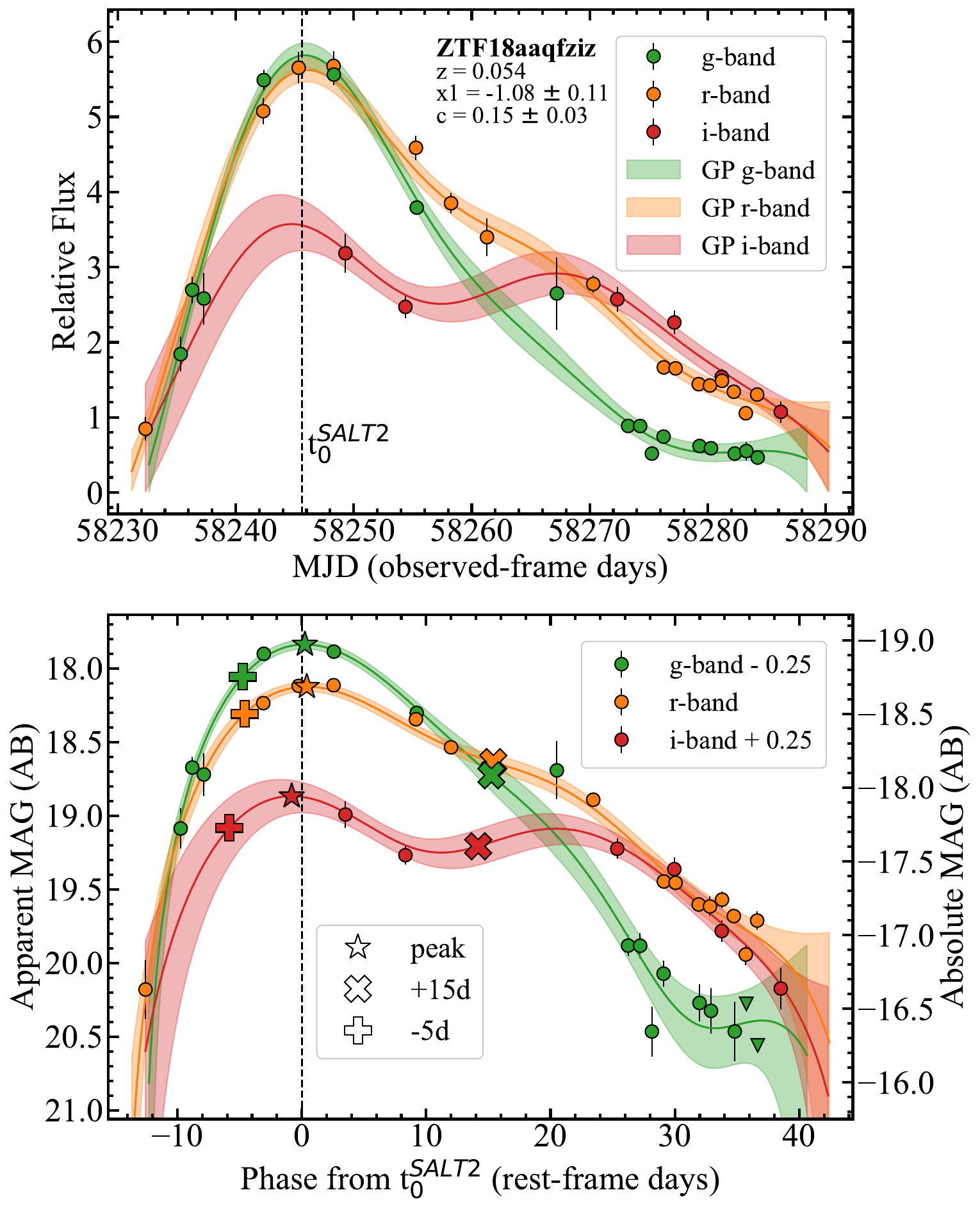}
    \caption{Example of a GPs fit to the light curves of ZTF18aaqfziz (SN\,2018bhp), a SN with average sampling (see Fig. 1 from \citealt{Rigault2024aAA}). \textbf{Top:} Light curves in flux space for $g$, $r$, and $i$-band, as a function of time (in observed days), colour-coded as described in the legend. The solid lines show the GP fits, as described in the text, with shaded regions displaying the $1\sigma$ uncertainty. The vertical dashed line marks the fitted \texttt{SALT2} time of maximum luminosity in $B$-band. \textbf{Bottom:} Same light curves and GP fits as above, but in magnitude space (both apparent and absolute), with downward triangles marking non-detections. The symbols correspond to the estimated brightness in each photometric band at various phases (see legends).}
    \label{fig:gp_fit}
\end{figure}

In the following, we briefly describe the main stages of the GPs fit method we use, and its application to the ZTF SN Ia DR2 sample, with a detailed outline presented in \citet{Deckers2024AA}. Firstly, for each SN, the flux data are averaged nightly (with the individual flux measurements' uncertainty as weight) and then corrected for the Milky Way extinction along the line of sight, assuming a \citet{Fitzpatrick1999PASP} reddening law with $R_{V}=3.1$. We then apply K-corrections based on the transient's redshift and sub-type, as discussed in Sect.~\ref{sec:kcor} and, finally, we fit the light curves. We use the python package \texttt{scikit-learn}\footnote{\url{https://scikit-learn.org/stable/modules/gaussian_process.html}} to perform the GP regression. For the GP object, we follow the prescription applied in \citet{Boone2019AJ} for \texttt{avocado}: We choose a mean function of zero (treating the SN light curve as a pure perturbation on a flat background) and the radial basis function (RBF) kernel (also known as the `squared exponential' kernel), which generally preserves the smoothness of the SN light curve evolution (in contrast with, e.g. the Mat\'ern kernels), with the addition of a constant and a white noise kernel (see \citealt{Deckers2024AA} for a discussion). Our fits are performed in flux space, so that non-detections are contributing in the regression, and in both time and wavelength space, combining information from all photometric bands simultaneously. The hyperparameters of the kernels are optimized by maximizing the log-marginal-likelihood, with the exception of the length scale in wavelength for which we fix it to 6,000 \AA. This 2D-fit scheme proves to be particularly helpful for non-homogeneous light curves, for instance in cases of missing data points in one of the photometric bands. This can be seen, for example, in Fig.~\ref{fig:gp_fit}, where there are no $i$-band observations at the early epochs of ZTF18aaqfziz (SN\,2018bhp); however, the GP fit generates a reasonable $i$-band light curve, although with higher uncertainty.

The product of this process is an \texttt{sklearn} class of the \texttt{GaussianProcessRegressor} for each SN, with which we are able to estimate various light curve parameters at the time frame that our GPs fits can provide: the time of maximum, the peak brightness (both in apparent and absolute magnitude, using the distance modulus obtained from the redshift) and the magnitude decline 15 days after maximum ($\Delta m_{15}$, or, in fact, relative changes of the light curve in any timescale) for each photometric band, the colours\footnote{The phase of the colours in this work are relative to the time of the bluer filter; for example $(g-r)_{\mathrm{peak}}$ refers to the colour at $g$-band peak brightness.} and the time difference between the maxima of two photometric bands. The estimated value for each requested parameter is the mean of the predictive distribution, with the standard deviation as the formal uncertainty. To estimate the uncertainties at the timings of the light-curve maxima, we follow \citet{Deckers2024AA} and perform an iterative resampling procedure (`bootstrapping') and fit. For light-curve parameters that are a combination of two or more, uncertainties are propagated and added in quadrature. Finally, we track the photometric coverage of each light curve by measuring the number of epochs (i.e. the nightly averaged photometric points) in various phase bins. A sub-set of the parameters we estimate is shown in Table~\ref{tab:samples_params} of Sect.~\ref{sec:sample_statistics} of the Appendix, while the complete set of parameters are available at \href{http://ztfcosmo.in2p3.fr}{ztfcosmo.in2p3.fr} and at the CDS.

\subsection{Final samples}
\label{sec:final_samples}

After applying the appropriate K-corrections and fitting the light curve data with GPs, we construct sub-samples of the population, in order to systematically study their light-curve diversity. These samples are primarily defined from the coverage of their light curve and their rarity: For the more common sub-types (normal, 91T-like and 91bg-like) we use consistent coverage conditions, while for the rest and less common sub-types, we additionally inspect their fits visually to determine whether they will be added to a sub-sample. We note that these sub-samples and their relative population numbers do not correspond to any physical rates, since, for this part of our analysis, we are mainly interested in the photometric properties of our samples.

We define two basic sub-samples, the `gr' and the `gri' sample: The gr sample corresponds to good photometric coverage in the $g$ and $r$ bands, and is defined as having at least one epoch in the $-3$ to $+3$ d and $+12$ to $+18$ d phase bins for both the $g$ and $r$ band, and at least one epoch at the $-10$ to $-3$ d and $+3$ to $+12$ d phase bins for either the $g$ or the $r$ band. The `gri' sample is a sub-set of the `gr' one, and introduces coverage conditions for the $i$-band, requiring at least one epoch at all the aforementioned phase bins. This stricter approach was preferred because of the more complex $i$-band's light curve evolution, due to the strong secondary maximum present at post-peak. Moreover, we employ three fit quality cuts: We dismiss events with uncertainties more than 3$\sigma$ away from the median of the uncertainty distribution for the peak magnitude ($g$ and $r$ band for gr, plus $i$ band for gri), the magnitude decline rate 15 days after peak ($\Delta m_{15,g}$ and $\Delta m_{15,r}$ for gr, plus $\Delta m_{15,i}$ for gri) and the peak magnitude time difference ($t_{max}^{r-g}$ for gr, plus $t_{max}^{i-g}$ for gri). Finally, and having no estimate of a possible host-galaxy extinction, we include an additional cut on the cosmological sub-types (normal and 91T-like) of $c<0.3$, in order to reject heavily extinct events. This removes 50 (from 1,143) normal and 7 (from 133) 91T-like events, that are primarily located close to the host galaxy ($d_{DLR}<1$ and $<0.6$, respectively), with no effect on our final results.

With these choices, we ensure good precision with large-number statistics for the gr sample and accurate precision, at the cost of numbers, for the gri sample, in tracking the photometric behaviour from $\sim$10 before to $\sim$25 rest-frame days after peak for each event. In order to avoid the Malmquist bias, we additionally consider the `gr-zcut' and `gri-zcut' samples, with a spectroscopic cut at $z\leq0.06$. The statistics of these sub-samples are shown in Table~\ref{tab:samples_numbers} of Sect.~\ref{sec:sample_statistics} of the Appendix.

\section{Results}
\label{sec:results}

In this section, we present the results of our light-curve modelling, focusing on the photometric properties of the peculiar thermonuclear population, and we investigate methods to identify them in samples of SNe Ia dominated by normal events. We additionally explore their host galaxy properties, in association with the normal population. Finally, we estimate the observed fractions of the SN Ia sub-classes in our volume-limited sample.

\subsection{Photometric properties}
\label{sec:results_phot_prop}

\begin{figure*}
    \centering
    \includegraphics[width=2\columnwidth]{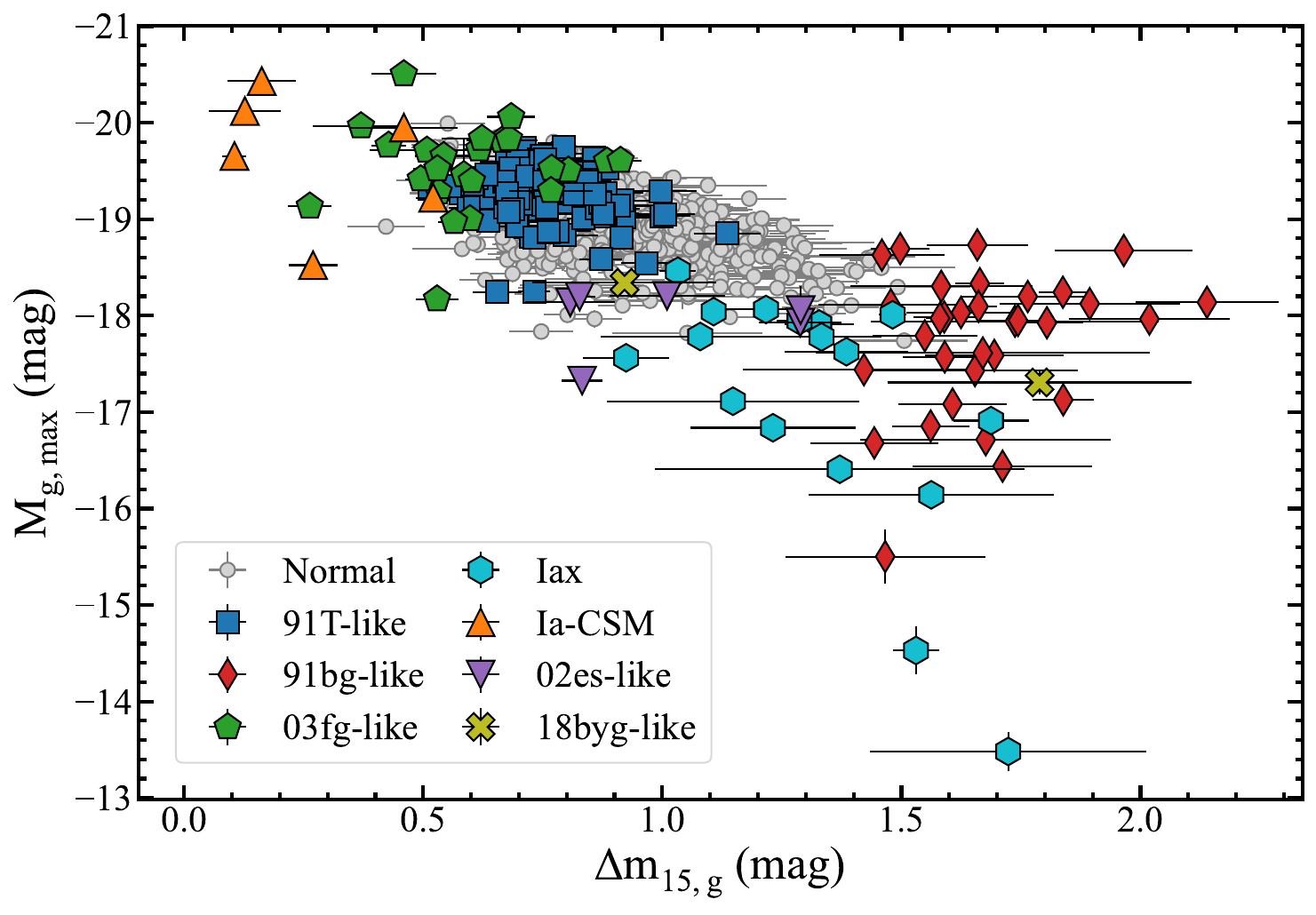}
    \caption{Absolute $g$-band magnitude at peak vs the decline rate within 15 days of peak in the $g$ band of the gr sample. The different sub-types are presented in different symbols and colours, as shown in the legend.}
    \label{fig:abs_mag_vs_decline_rate}
\end{figure*}

In Fig.~\ref{fig:abs_mag_vs_decline_rate}, we show the results of our GP fits to the gr sample in the light curve luminosity versus width plane, probed by the absolute magnitude at the $g$-band peak, $M_\mathrm{{g,max}}$ and the $g$-band decline rate $\Delta m_\mathrm{{15,g}}$, respectively. This is the largest homogeneous SN Ia sample showcasing the diversity of the thermonuclear population to date.

We estimate a median absolute peak magnitude in $g$ band for the gr-zcut normal SNe Ia of $-18.93$ with a standard deviation of $0.34$ mag and a decline rate of $0.88\pm0.18$ mag, respectively. At the same time, they show the well-established brighter-slower correlation \citep{Phillips1993ApJ}, as the slower evolved SNe (with a decline rate lower than the median) are $\sim0.27$ mag brighter than the faster ones.

\begin{figure}
    \centering
    \includegraphics[width=1\columnwidth]{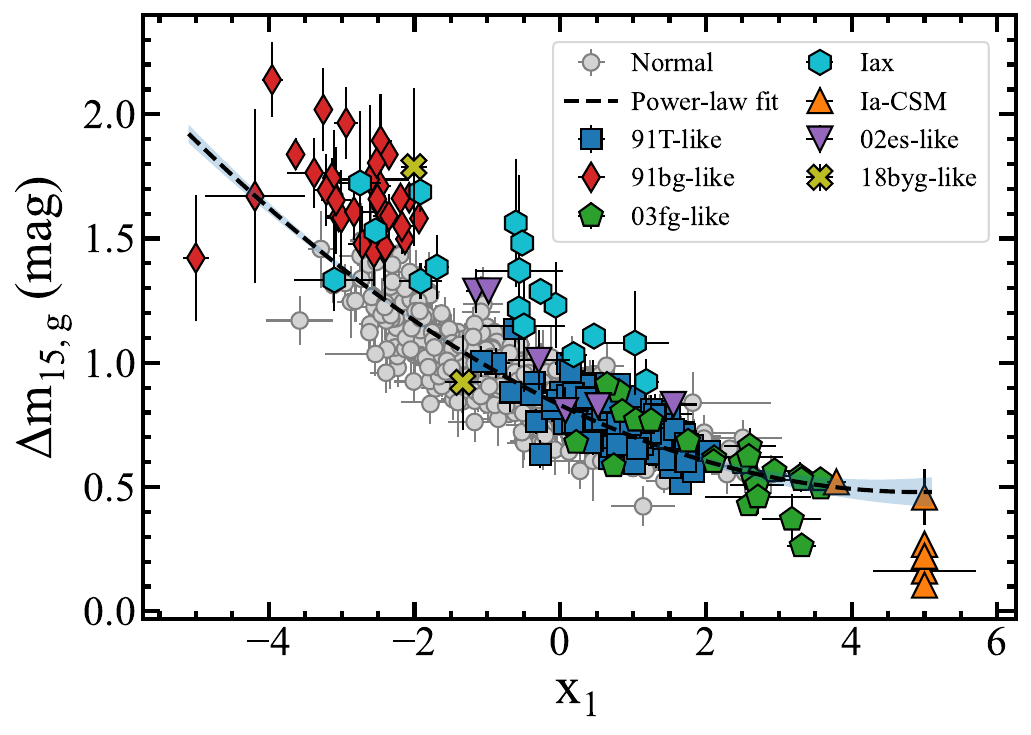}
    \caption{\texttt{SALT2} $x_{1}$ light curve stretch parameter vs the decline rate $\mathrm{\Delta m_{15,g}}$ for the gr sample, as described in the legend. The solid dashed line corresponds to a second-order power-law fit to the normal SNe Ia.}
    \label{fig:x1_vs_dm15}
\end{figure}

Previous studies have derived relationships between various stretch parameters and the $B$-band decline rate \citep{Guy2007AA,Conley2008ApJ,Siebert2019MNRAS}. However, as future optical surveys will observe in SDSS-like photometric bands ($ugri$ instead of Bessel/Johnson $BVRI$), a similar relationship with the $g$-band decline rate will be useful. Fig.~\ref{fig:x1_vs_dm15} shows our estimated $\mathrm{\Delta m_{15,g}}$ against the \texttt{SALT2} $x_{1}$ stretch for the gr sample. We fit the normal SN Ia data with a second order power-law and we derive an equation the can be used to convert \texttt{SALT2} fits to a decline rate in the $g$ band in future studies:

\begin{equation}
\label{eq:x1_to_dm15}
      \Delta m_{\mathrm{{15,g}}} = 0.014(0.002)x_{1}^{2} - 0.141(0.002)x_{1} + 0.830(0.002) \nonumber
\end{equation}

\begin{figure*}
    \centering
    \includegraphics[width=2\columnwidth]{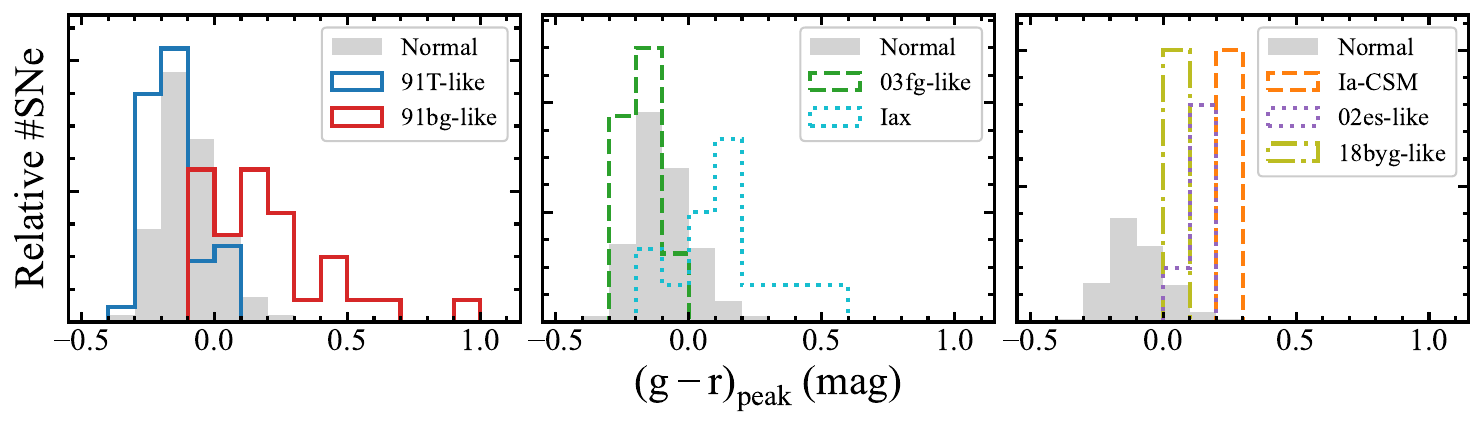}
    \caption{The gr-zcut sample's $g-r$ colour distribution at peak $g$-band brightness of the normal SNe Ia (in grey), compared with the 91T- and 91bg-like (\textbf{left}), 03fg-like and Iax (\textbf{middle}), and Ia-CSM and 02es-like (\textbf{right}) populations.}
    \label{fig:gr_color_peak}
\end{figure*}

Fig.~\ref{fig:gr_color_peak} shows distributions of the $g-r$ colour at peak for the gr-zcut normal SNe Ia sample, compared with the same distributions of the other sub-types. The normal SN Ia distribution has a median of $-0.11$ mag and a standard deviation of $0.10$ mag. We note that none of the SNe is corrected for possible host galaxy extinction on the line-of-sight. Our $c<0.3$ cut simply removes the very reddened normal/91T-like events (which probably are indeed affected by extinction in the host galaxy) but even within this clean sample, there is a colour diversity which is a combination of the intrinsic colour diversity and dust reddening (see \citealt{Ginolin2024AA} and \citealt{Popovic2024AA} for a further discussion on the ZTF SN Ia DR2 sample).

In the following sub-sections, we discuss the photometric properties of the peculiar sub-classes in detail.

\subsubsection{91T- and 91bg-like SNe Ia}
\label{sec:results_phot_prop_91t_91bg}

For the peculiar sub-class of the overluminous 91T-like, and focusing on the gr-zcut sample, we find a median absolute $g$-band peak magnitude of $-19.20\pm0.28$ and a decline rate of $0.72\pm0.09$. 91T-like SNe are generally brighter than the normal SNe Ia ($\sim0.25$ mag brighter than the median) and slightly brighter ($\sim0.15$ mag) than the bright end of the normal population. Their decline rate is $\sim0.16$ mag slower than the median of normal SNe Ia and their $g-r$ colour at peak is $-0.17\pm0.10$, $\sim 0.06$ mag bluer than the median of normal SNe Ia. Moreover, we do not find any significant difference in peak brightness between the 91T- and 99aa-like SNe, with the 99aa-like population having slightly faster decline rates and less blue colours (similar to the colours of the bright end of the normal SNe Ia). It appears that, from a photometric point of view, 91T-like SNe Ia are indistinguishable from the brighter normal SNe Ia, making their peculiarity strictly spectroscopic.

On the other hand, the peculiarities of the underluminous 91bg-like SNe Ia are both spectroscopic and photometric. This sub-class (for the gr-zcut sample) has a median absolute $g$-band peak magnitude of $-17.95\pm0.69$ and a decline rate of $1.66\pm0.17$ mag, substantially different from the normal SNe Ia, and with a wider range of photometric parameters. Moreover, they are substantially redder, with peak colours of $0.14\pm0.24$. We note that the separation of 91bg-like from the normal SNe Ia is also evident if we use the fits from the \texttt{SALT2} K-corrected light curves, for which 91bg-like become $\sim0.17$ mag fainter, $\sim0.22$ mag slower, and $\sim0.12$ mag redder, compared to the template K-corrected light curves.

\subsubsection{03fg-like SNe Ia}
\label{sec:results_phot_prop_03fg}

Apart from the traditional peculiar sub-classes of 91T- and 91bg-like SNe Ia, the sub-type with the most members in ZTF SN Ia DR2 is the 03fg-like SNe Ia. This type of explosions are relatively rare, thus they tend to be found in higher redshifts, posing difficulties for a direct statistical comparison of their properties with low-redshift SNe Ia. In our sample (Tables~\ref{tab:class_numbers} and \ref{tab:samples_numbers}), 38\% of 03fg-like SNe are at $z\leq0.06$, with their gr-zcut consisting only 8 members. Acknowledging this Malmquist bias, and for the full gr sample, 03fg-like SNe Ia have a median absolute $g$-band peak magnitude of $-19.56\pm0.44$, a decline rate of $0.59\pm0.15$ and $g-r$ colour at peak of $-0.15\pm0.08$ mag (note that for this and the following sub-classes, no cut in the \texttt{SALT2} $c$ parameter and no correction for possible host extinction has been applied). The brightest event in the sub-class is ZTF20abbbumr (SN\,2020kvl) with $-20.5\pm0.04$ mag, and the slowest is ZTF19aailltc (SN\,2019avn) with a decline rate of $0.26\pm0.05$ mag, reaching parameter values seen in SNe Ia-CSM.

Of particular interest is the faintest event, ZTF20ackitai (SN\,2020xqb): its peak magnitude ($-18.17\pm0.06$ mag) is low, even compared with the normal SNe Ia sample, and its $g-r$ colour at peak is $-0.06\pm0.03$ mag; it is one of the reddest 03fg-like events and shows no evidence of host-galaxy extinction. ZTF20ackitai appears similar to ASASSN-15hy \citep{Lu2021ApJ}, occupying a space of faint and slow thermonuclear events previously thought to be void. Finally, some 03fg-like events appear to have similar brightness and decline rates with the normal SNe Ia population (Fig.~\ref{fig:abs_mag_vs_decline_rate}). These five events, ZTF19aanmdsr (SN\,2019cng), ZTF19abnimpq (SN\,2019myl), ZTF20aamhocj (SN\,2020bzg), ZTF20acnyxln (SN\,2020yjf) and ZTF20acyybvq (SN\,2020adhz) have  $\mathrm{\Delta m_{15,g}}>0.75$ mag and resemble SN\,2012dn \citep{Chakradhari2014MNRAS,Taubenberger2019MNRAS}, hinting for an internal diversity in the sub-class and the possibility for a continuum between 03fg-like and normal SNe Ia \citep{Chen2019ApJ,FitzAxen2023ApJ}. 

\subsubsection{SNe Iax}
\label{sec:results_phot_prop_iax}

SNe Iax is the most common peculiar sub-type at low redshifts, with 65\% of the events in the gr-zcut sample. SNe Iax show a large range in the brightness and decline rate space, with members of the sub-class being as luminous as the fainter normal SNe Ia and as faint as the most energetic Luminous Blue Variables (LBVs). Following \citet{Srivastav2022MNRAS}, we split the population into `bright' and `faint' SNe Iax at mag$ =-16$, and for the bright population, we find a median absolute $g$-band peak magnitude of $-17.78\pm0.65$, a decline rate of $1.29\pm0.20$ and $g-r$ colour at peak of $0.16\pm0.18$ mag. The faint population includes the two well-observed events, ZTF19aawhlcn \citep[SN\,2019gsc;][]{Tomasella2020MNRAS,Srivastav2020ApJ} and ZTF20abaunmw \citep[SN\,2020kyg;][]{Srivastav2022MNRAS}, with peak magnitudes of $-13.48\pm0.20$ and $-14.53\pm0.25$ mag, respectively.

It has been suggested that a linear correlation for the peak brightness and the decline rate of SNe Iax do exist, with brighter events having longer timescale evolution, probed either by $\mathrm{\Delta m_{15}}$ \citep{Foley2013ApJ} or the rise time \citep{Magee2016AA}. We calculate the Pearson correlation coefficient for the decline rate versus peak absolute magnitude in $g$ and $r$ band, and we find a statistically significant correlation coefficient of 0.657 ($p$-value of $\sim$0.004) and 0.685 ($p$-value of $\sim$0.002), respectively. As we do not have accurate estimates of the rise times, we instead calculate the Pearson correlation coefficient for the rise rate $\Delta m\mathrm{_{-5}}$ (the magnitude rise 5 days before maximum) versus peak absolute magnitude in $g$ and $r$ band, for which we find similar values, suggesting that a correlation between peak brightness and light curve timescale evolution does exist.

\subsubsection{SNe Ia-CSM}
\label{sec:results_phot_prop_iacsm}

While easily identifiable due to their distinct spectroscopic properties (particularly the presence of strong H$\alpha$), SNe Ia-CSM remain one of the rarest thermonuclear explosions. Excluding the He-rich SN Ia-CSM ZTF20aatxryt \citep{Kool2023Natur}, the SNe Ia-CSM in our sample are almost identical with the ones presented in \citet{Sharma2023ApJ}, and generally show bright and slow light curves, with median absolute $g$-band peak magnitudes of $-19.80\pm0.63$ and a decline rate of $0.22\pm0.16$ mag. SNe Ia-CSM at peak are redder than normal SNe Ia ($g-r=0.20\pm0.13$ mag), attributed to the interaction of the SN ejecta with the H-rich CSM, resulting in strong H$\alpha$ emission.

\subsubsection{02es-like SNe Ia}
\label{sec:results_phot_prop_02es}

The 02es-like SNe Ia represents a class of objects with light curve evolution timescales similar to normal SNe Ia, but substantially fainter, and with spectroscopic similarities with the 91bg-like SNe Ia. The 02es-like sample in ZTF SN Ia DR2 has a median absolute $g$-band peak magnitude of $-18.10\pm0.31$, a decline rate of $0.92\pm0.21$ mag and $g-r$ color at peak of $0.14\pm0.03$ mag. The peculiar ZTF19adcecwu \citep[SN\,2019yvq;][]{Miller2020ApJ} has a similar brightness with the other events, but with a faster decline rate ($1.29\pm0.02$ mag), although still in the normal SNe Ia region. We additionally identify one event, ZTF18abmjyvo (SN 2018fju), typed as an 02es-like, with similar photometric properties as ZTF19adcecwu.

\subsubsection{18byg-like SNe Ia}
\label{sec:results_phot_prop_18byg}

The defining characteristic of the 18byg-like sub-type is the strong flux suppression at the bluer part of the spectrum, originally attributed to line blanketing from Fe-group elements. We have identified 3 events, apart from the original 18byg-like (ZTF18aaqeasu), ZTF19abttrte (SN\,2019otv), ZTF20aayhacx (SN\,2020jgb) and ZTF20achoqvb (SN\,2020xvf) that show a similar characteristic feature at shorter wavelengths. Unfortunately, only one of them (ZTF20aayhacx) has sufficiently good photometric coverage to be included in our gr sample, thus our results are very limited due to small number statistics.

For both of the events in our sample, we find relatively low $g$-band peak magnitudes of $-17.31\pm0.12$ and $-18.34\pm0.08$ (compared to normal SNe Ia), with no signs of host extinction in their spectra, such as sodium absorption lines at the host's redshift. However, the decline rates vary substantially, with ZTF18aaqeasu having 91bg-like $\mathrm{\Delta m_{15,g}}$ and ZTF20aayhacx declining as a normal SNe Ia. Moreover, ZTF20aayhacx has a $g-r$ color at peak of $0.02\pm0.05$ mag, placing it at the redder part of normal SNe Ia, but certainly within their distribution, while ZTF18aaqeasu is considerably redder ($g-r = 0.78\pm0.12$ mag). We also note that for these two events, the \ion{Si}{ii}$ \lambda5972$ over \ion{Si}{ii}$ \lambda6355$ pEW ratio is also different, with ZTF18aaqeasu having a pronounced \ion{Si}{ii}$ \lambda5972$ line while for ZTF20aayhacx, the line is very weak. Finally, ZTF20aayhacx shows a hint of a flux excess (seen only in $g$ band, due to limited coverage) at $\sim$14 days before peak, resembling the strong flux excess of ZTF18aaqeasu $\sim$15-17 days before maximum.

\subsection{Distinguishing between sub-types from photometry}
\label{sec:results_time_vs_color15}

The two main distinguishing light curve features of normal SNe Ia are the relative timing of the maxima at bluer versus redder wavelengths and the presence of a prominent secondary maximum $\sim$15--30 days after the primary maximum in their $i$-band and NIR light curves (see \citealt{Deckers2024AA} for a comprehensive study of the secondary maximum in the ZTF SN Ia DR2 sample). The secondary maximum is thought to originate from the ionization evolution of iron group elements in the ejecta: iron and cobalt have an increased emissivity in the NIR at temperatures $\sim$7,000 K due to the transitioning from double to single ionised state \citep{Kasen2006ApJ}. The onset and strength of the secondary maximum correlates with the $^{56}$Ni synthesised in the explosion, with high $^{56}$Ni yields leading to brighter explosions, higher temperatures, sustained high opacity for a longer time and more pronounced secondary maxima. However, effects such as the total ejecta mass, the mixing of $^{56}$Ni to the outward layers, the metallicity of the progenitor star or even explosion mechanisms different from the canonical model (such as models with an additional heating source apart from radioactivity) may alter the standard behaviour. 

\begin{figure}
    \centering
    \includegraphics[width=1\columnwidth]{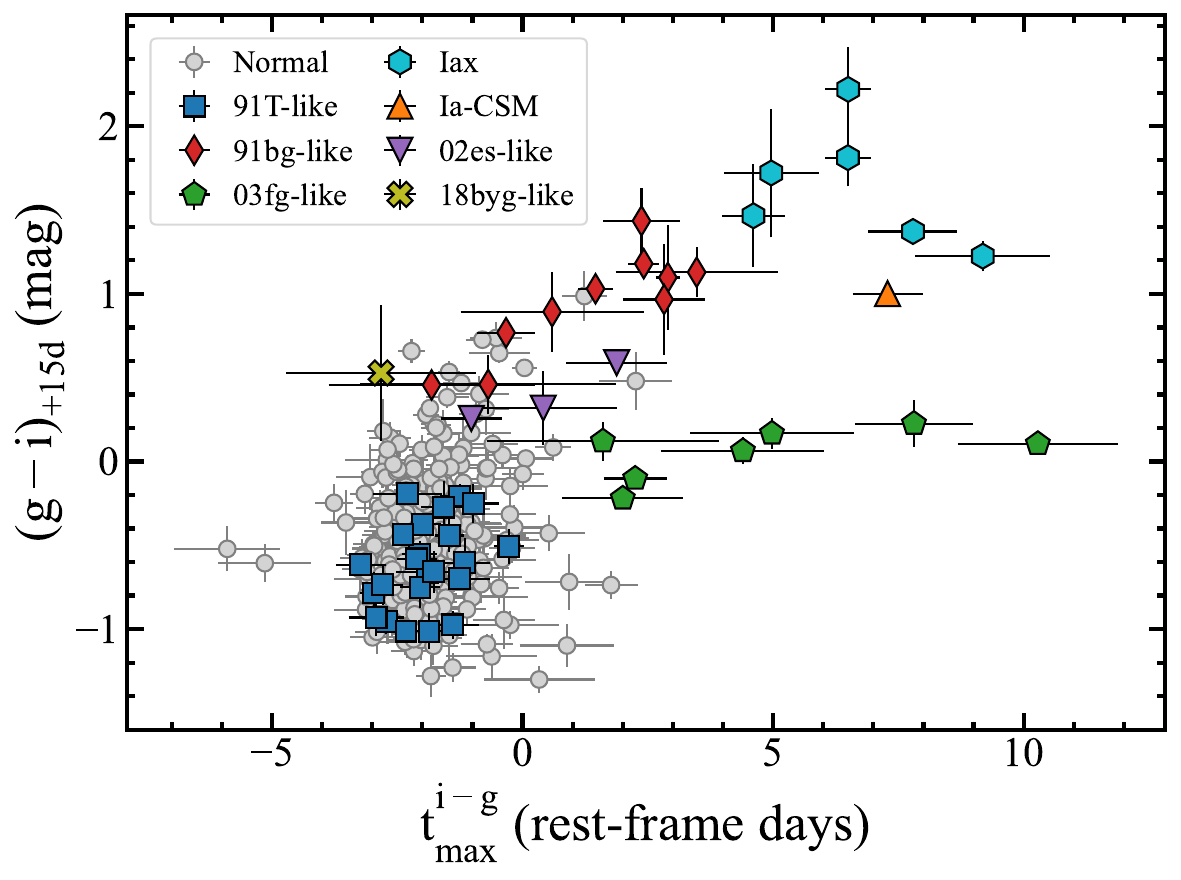}
    \caption{$g-i$ colour at $+15$ days from the peak $g$-band magnitude as a function of the time difference (in rest-frame days) between the primary maxima in the $i$ and $g$ bands of the gri sample. The different sub-types are presented in different symbols and colours, as shown in the legend.}
    \label{fig:color_ig_vs_t_ig}
\end{figure}

Motivated by the work of \citet{Ashall2020ApJ}, where they used the \texttt{SNooPy} \citep{Burns2011AJ} stretch parameter $s_{BV}$ and the time difference of the $B$- and $V$-band peak, we explore two parameters calculated from our GP fits, the time difference of the peak of a red versus a blue photometric band and the colour of the same photometric bands at $+15$ days from maximum. These two parameters encapsulate the ionisation state of the event at peak and the strength of the NIR secondary maximum. 

Firstly, we focus on the behaviour of the $g$- and $i$-band light curves, corresponding to the bluest and reddest photometric bands of ZTF. Fig.~\ref{fig:color_ig_vs_t_ig} shows the $g-i$ color at $+15$ days from maximum with respect to the time difference $t_{max}^{i-g}$ of the $i$- and $g$-bands primary maxima of the gri sample. The normal SNe Ia have a median $t_{max}^{i-g}$ of $-1.95$ and a standard deviation of $0.94$ days, meaning that the $i$-band light curve peaks prior to the $g$-band one. Their $g-i$ colour is $-0.51\pm0.40$, with bluer colours corresponding to brighter events. The 91T-like SNe Ia completely overlap with the luminous normal SNe Ia population, with a $t_{max}^{i-g}$ median of $-1.99$ d and a standard deviation of $0.71$ d and a $g-i$ colour of $-0.62\pm0.25$ mag.

The remaining sub-classes are clearly separated. The 91bg-like SNe Ia show a red $g-i$ colour $(1.00\pm0.29$ mag) and a positive $t_{max}^{i-g}$ of $1.91\pm1.70$ d, with redder events having larger $t_{max}^{i-g}$, forming possibly a continuum with the fainter normal SNe Ia. 02es-like SNe Ia have a similar behaviour, but with slightly bluer colours. For the prototypical 18byg-like event, we find it to be placed at negative $t_{max}^{i-g}$ and relatively red $g-i$ colour; however, we note the large uncertainty on the estimated parameters. The 03fg-like SNe Ia have a large spread in $t_{max}^{i-g}$ with $4.40\pm3.02$ d while their $g-i$ colour shows little variation with $0.11\pm0.15$ mag. For the one SN Ia-CSM with good $i$-band coverage, ZTF18aaykjei (SN\,2018crl)\footnote{Note that the redshift of ZTF18aaykjei is $z \sim 0.097$ and no k-corrections have been applied to SNe Ia-CSM.}, we estimate $t_{max}^{i-g}=7.28\pm0.70$ d and $g-i=1.00\pm0.05$ mag. Finally, SNe Iax have the redder $g-i$ colours, with $g-i=1.59\pm0.33$ mag and the largest $t_{max}^{i-g}$, with $t_{max}^{i-g}=6.50\pm1.57$ d.

\begin{figure}
    \centering
    \includegraphics[width=1\columnwidth]{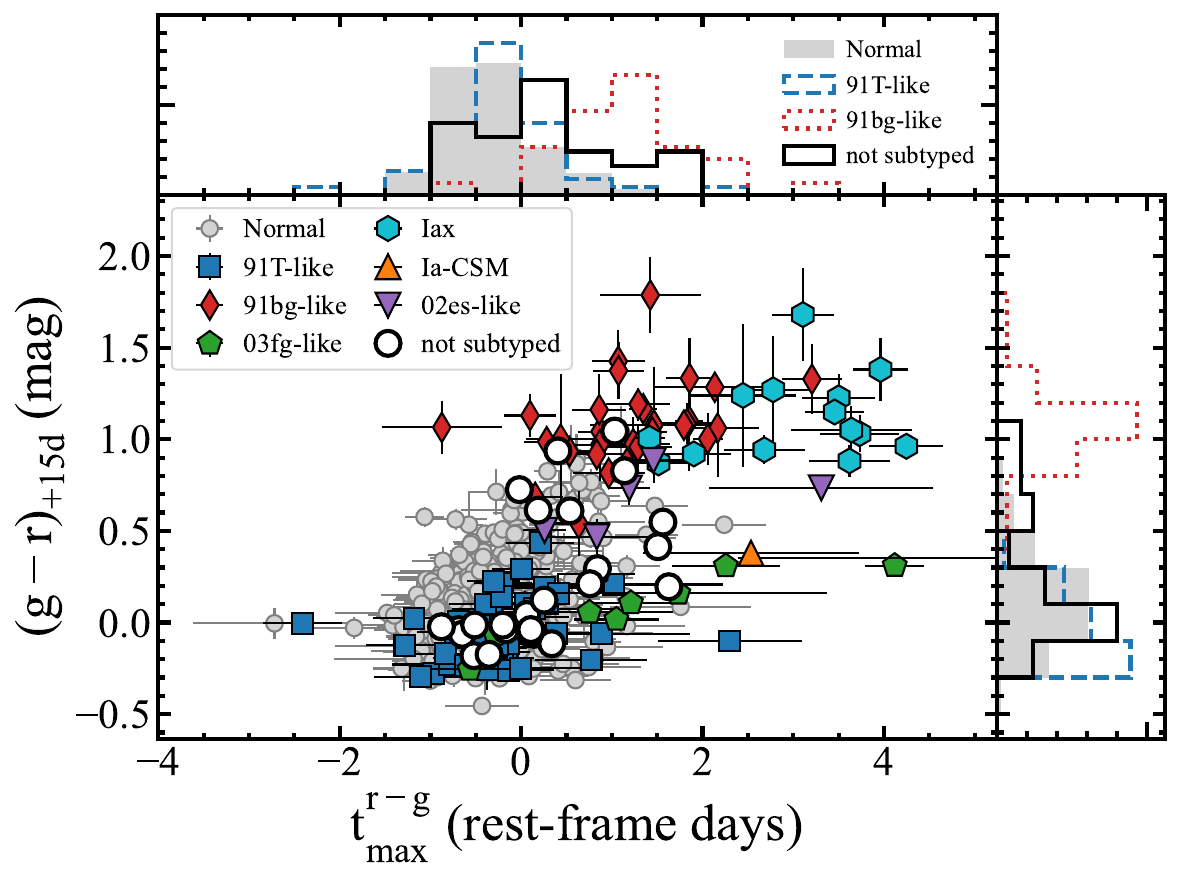}
    \caption{$g-r$ colour at $+15$ days from the peak $g$-band magnitude as a function of the time difference (in rest-frame days) between the primary maxima in the $r$ and $g$ bands of the gr-zcut sample. The different sub-types are presented in different symbols and colours, as shown in the legend. We additionally show the distribution of the parameters for the normal, 91T-like, and 91bg-like SNe Ia, alongside the not-sub-typed events, as shown in the legend.}
    \label{fig:color_rg_vs_t_rg}
\end{figure}

As the ZTF coverage in the $r$ band was far superior than the one in the $i$ band, resulting in more events in the gr relative to the gri sample, we perform a similar analysis, but now using the $g-r$ colour at $+15$ days and $t_{max}^{r-g}$. We restrict ourselves to the gr-zcut sample, but we now also include the not sub-typed events of the sample, in an effort to provide a classification. Our results are shown in Fig.~\ref{fig:color_rg_vs_t_rg}. The regions that the various sub-types occupy resemble the ones in Fig.~\ref{fig:color_ig_vs_t_ig}, but with more scatter. This is expected, as the NIR emissivity of the $r$-band is substantially lower, compared to the $i$-band, with the light curves forming a `shoulder' instead of a secondary maximum, affecting the colour distinction between the sub-types.

We do not find any not sub-typed event to overlap with the regions occupied by the peculiar sub-classes. The majority of them lie in the normal/91T-like region, confirming our suspicion that these events are either normal or 91T-like, but the classifiers were unable to distinguish (see Sect.~\ref{sec:subclassification} for a discussion). In particular, the not sub-typed events with $(g-r)_{\mathrm{15d}}>0.3$ are mainly SNe that suffer from high host galaxy extinction ($c>0.4$ for this subsample), reducing the quality of the spectrum due to the contamination of the host's light in the slit. On the other hand, 14 of the 21 SNe with $(g-r)_{\mathrm{15d}}<0.3$ had spectra taken $>+15$ days from maximum, making the sub-classification difficult, particularly the distinction between a normal and a 91T-like SN Ia (see \citealt{Burgaz2024AA} for further discussion).Nevertheless, we use this rough sub-typing in the calculation of the peculiar events relative rate.

\subsection{The host galaxies of the peculiar SNe Ia}
\label{sec:results_hosts}

\begin{figure*}
    \centering
    \includegraphics[width=2\columnwidth]{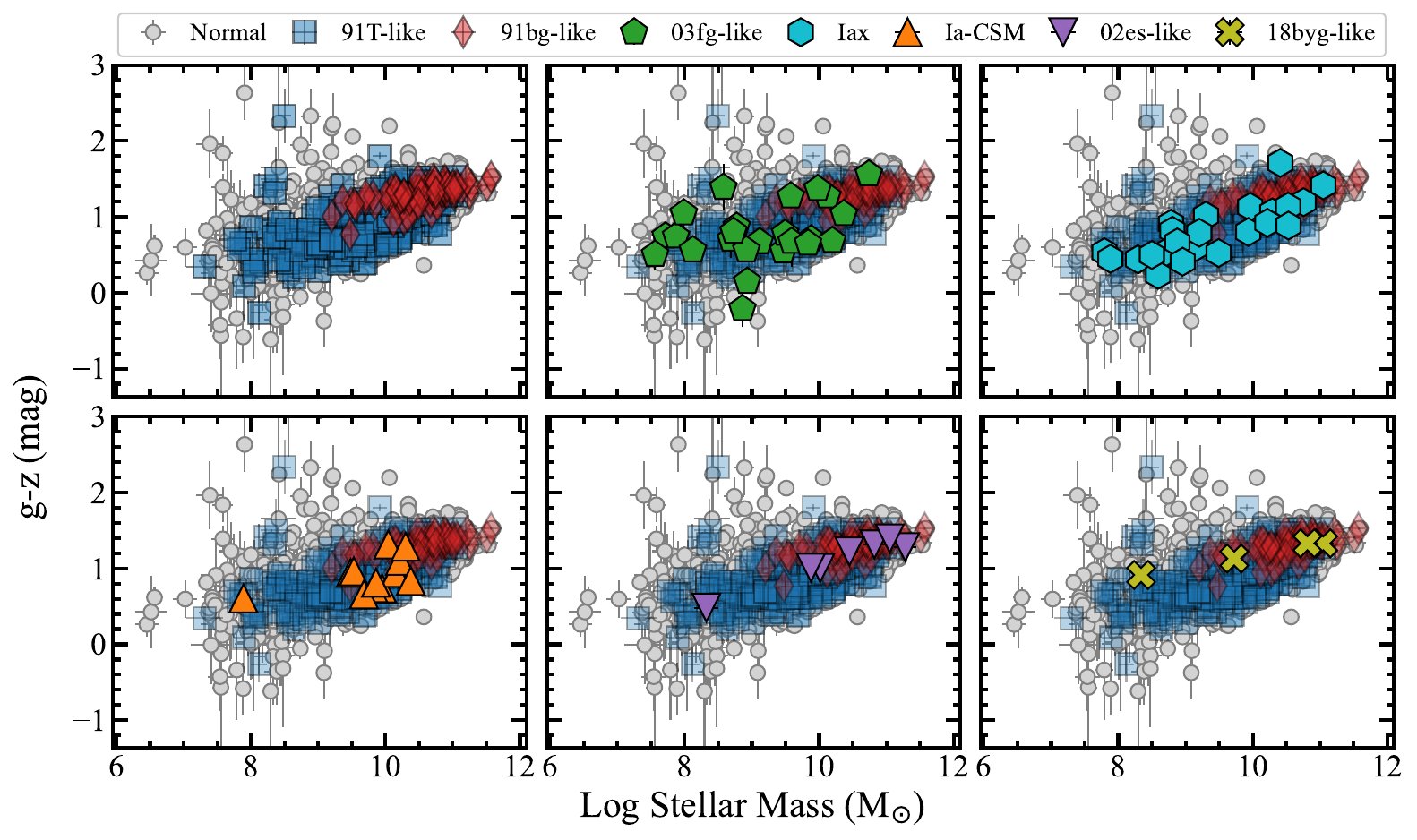}
    \caption{Rest-frame $g-z$ host galaxy colour vs. the logarithm of the host galaxy stellar mass, as calculated in \citet{smith2024AA}. The different sub-types are presented in different symbols and colours, as shown in the legend. All host galaxies in the plots have spectroscopic redshifts (see Sect.~\ref{sec:metadata}).}
    \label{fig:hosts_color_mass}
\end{figure*}

One of the most promising ways to associate a particular progenitor scenario with a class of objects is by studying their host galaxy properties. In Fig.~\ref{fig:hosts_color_mass}, we compare the host galaxy masses (in units of logarithm of solar mass $\mathrm{M_{\odot}}$) and rest-frame $g-z$ colours (in units of magnitudes, roughly tracking the (recent) specific star formation) of the normal SNe Ia with the peculiar sub-classes.

The normal SNe Ia host galaxies span the whole range of the mass-colour parameter space, with medians/standard deviations of $10.01\pm0.83$ dex and $1.01\pm0.35$ mag, respectively. The same trend is seen for the hosts of 91T-like SNe Ia, for which we estimate slightly lower masses of $9.85\pm0.78$ dex and bluer $g-z$ colours of $0.89\pm0.31$ mag. On the other hand, the hosts of 91bg-like SNe Ia have masses of $10.61\pm0.47$ dex and colours of $1.33\pm0.13$ mag, with a two-dimensional (2D) Kolmogorov-Smirnov (KS) test on the 91bg-like and normal populations having p-values of $1.4\times10^{-17}$, indicating that the two samples are different. We note that, while the 91bg-like SNe Ia are restricted to higher mass and redder environments, the 91T-like host galaxies are generally similar to the normal ones.

We now focus on the rarer peculiar sub-classes. The 03fg-like host galaxies are generally less massive and bluer than normal SNe Ia, with values of $9.02\pm0.87$ dex and $0.72\pm0.39$ mag, respectively, with a 2d KS test with the normal population giving a p-value of $0.0006$, indicating that these host galaxies may originate from different distributions. Interestingly, while most of the host galaxies of 03fg-like SNe Ia are low-mass and blue, a trend identified in the past for the 03fg-like events, we find that a sub-set of this population explodes in higher mass hosts with redder colours. For the host galaxies of Iax, we find masses of $9.20\pm0.92$ dex and colours of $0.80\pm0.35$ mag, with a 2D KS test with the normal population giving p-value of $0.0159$, indicating different underlying populations.

Finally, for the 02es- and 18byg-like events, although the numbers of the events are low, there are indications for similarities with the 91bg-like hosts, with p-values of $0.35$ and $0.15$, respectively. However, we note that both of these sub-classes have members with low-mass/blue colour host galaxies. ZTF20abrjmgi, which we tentatively classified as an 02es-like SN, is at the extreme low-mass side of the population. The same is true for the 18byg-like ZTF20aayhacx \citep{Liu2023ApJ}, but also for other 18byg-like events that are not in our sample, such as ZTF22aajijjf \citep[SN\,2022joj;][]{Liu2023ApJ2} and ZTF19aatheus \citep[SN\,2019eix;][]{Padilla2023ApJ}. On the other hand, the host galaxies of the SNe Ia-CSM appear to populate a narrow region in the host mass distribution around $\sim9.95\pm0.29$ dex, with the exception of the only He-rich Ia-CSM, ZTF20aatxryt, for which its host is a blue low-mass galaxy.

\begin{figure}
    \centering
    \includegraphics[width=1\columnwidth]{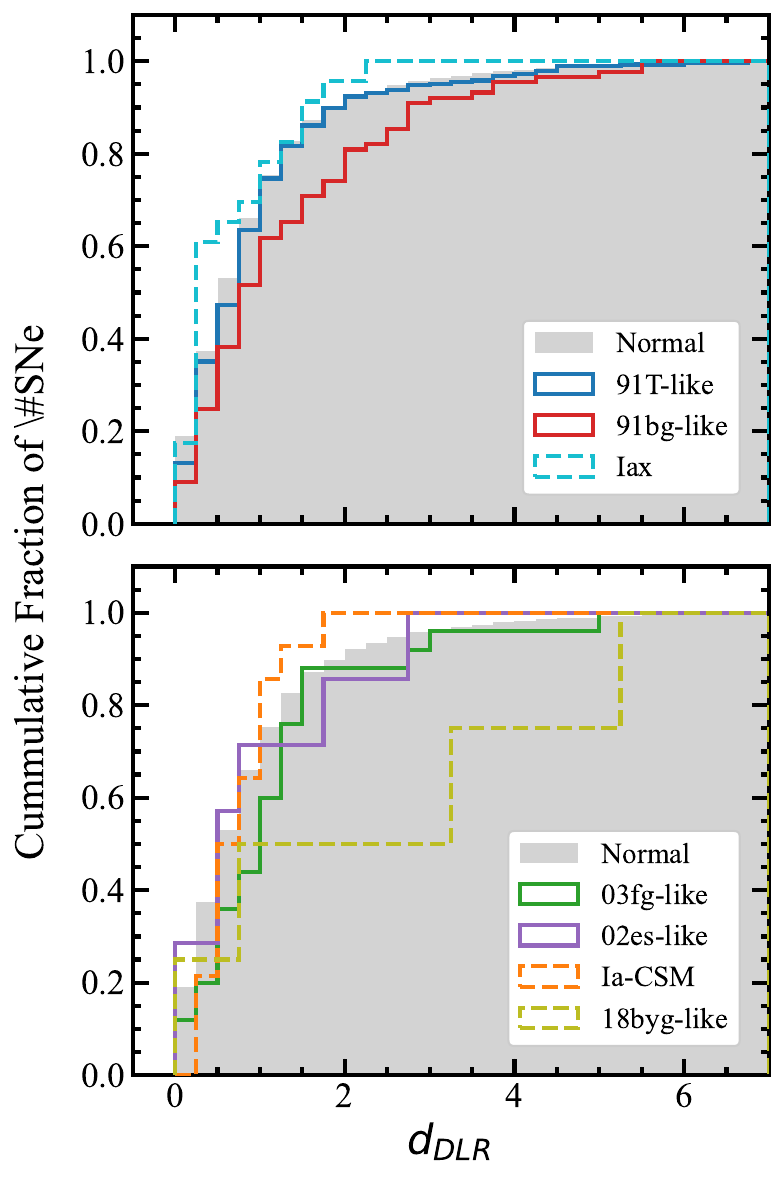}
    \caption{Cumulative distributions of the $d_{DLR}$ of the SNe in ZTF SN Ia DR2. The different sub-types are presented in different colours, as shown in the legends.}
    \label{fig:hosts_dDLR}
\end{figure}

Fig.~\ref{fig:hosts_dDLR} shows the cumulative distributions of the $d_{DLR}$ of the SNe in our sample, split into the different sub-types. The parameter $d_{DLR}$ is a proxy of the distance of the SN relative to its host, independent of its redshift\footnote{However, for the faint and/or high-redshift galaxies, the uncertainty in the estimated radius is increasing.}. Our results show that the distribution of the 91T-like $d_{DLR}$ are identical with the ones of normal SNe Ia, while the $d_{DLR}$ of the 91bg-like SNe originate from a different distribution (p-value of 0.002), generally preferring more remote locations. For the rarer sub-types, and acknowledging the low number statistics, we find a preference for remote locations for the 18byg-like compared to the other SN Ia sub-classes, and locations near the host-galaxy cores for the Iax and Ia-CSM SNe Ia, compared to normal SNe Ia. Interestingly, the 03fg- and 02es-like SNe Ia, although exploding in different galaxies (in terms of global mass and colour, see Fig.~\ref{fig:hosts_color_mass}) the distribution of their $d_{DLR}$ appears similar (p-value of 0.41).

\subsection{The relative rates of the SN Ia sub-classes}
\label{sec:results_rel_rates}

In this section, we estimate the relative population size for each of the sub-classes in the ZTF SN Ia DR2 sample. We do not attempt to calculate an absolute volumetric rate for each sub-type, but restrict ourselves to observed fractions.

\citet{Amenouche2024AA} have shown that the ZTF SN Ia DR2 sample is volume limited for $z\leq0.06$, but this refers to the normal SN Ia population. This means that, for events with intrinsic brightness lower than normal SNe Ia (such as Iax, 91bg-like, 02es-like, and 18byg-like), a correction must be applied due to the lower probability of detecting such transient. These corrections are usually estimated with simulations that take into account the survey observing schema, which includes the cadence, the observing conditions (effects of the weather for each observing night, technical difficulties or other), seasonal gaps, the survey footprint, and its magnitude limit. In order to estimate this efficiency, we use \texttt{skysurvey}\footnote{\url{https://skysurvey.readthedocs.io/en/latest/}}, a python package designed to simulate a survey output for sets of transient physical models.

We start by simulating a total of 6,840,000 transients, with their light curves generated from a model based on \texttt{SALT2} and the SN Ia rate calculated by \citet{Perley2020ApJ}. Each transient has a set of random \texttt{SALT2} $x_{1}$, $c$, and $t_{0}$, for which we use uniform priors of:

\begin{enumerate}
    \item $-5.0 \leq x_{1} \leq +5.0$
    \item $-0.4 \leq c \leq 1.1$
    \item $-21.0 \leq M\mathrm{_{B,max}} \leq -13.0$
    \item $58,200 \leq t_{0} \leq 59,200$ (the duration of the ZTF SN Ia DR2 survey, in MJD)
    \item $\mathrm{0^{\circ}\leq \mathrm{R.A.} \leq360^{\circ}}$ and $\mathrm{-30^{\circ}\leq \mathrm{Dec.} \leq+90^{\circ}}$ (the ZTF survey footprint)
\end{enumerate}

Then, \texttt{skysurvey} generates observed ZTF light curves based on the ZTF observing logs. A specific transient is considered detected if its observed light curve passes the criteria resulting in having `lccoverage\_flag=1' (see \citealt{smith2024AA} for more details). These criteria are applied to the phase range of -10 to 40 (rest-frame) days from $t_{0}$, are required to be satisfied simultaneously, and are defined as:

\begin{enumerate}
    \item The transient has detections\footnote{A detection is determined at the nightly stacked photometric point and at the $5\sigma$ level} in at least 2 filters.
    \item The transient has a detection in at least 1 filters pre-peak.
    \item The transient has a detection in at least 1 filters post-peak.
    \item The transient has at least 2 detections pre-peak.
    \item The transient has at least 2 detections post-peak.
    \item The transient has at least 7 detections across all filters.
\end{enumerate}

Finally, we estimate the efficiency $eff$ (i.e. the number of detected events divided with the number of simulated events) as a function of $x_{1}$, $c$, and peak absolute magnitude in bins with width of $0.25$, $0.025$, and $0.2$ mag, respectively.

\begin{figure*}
    \centering
    \includegraphics[width=2\columnwidth]{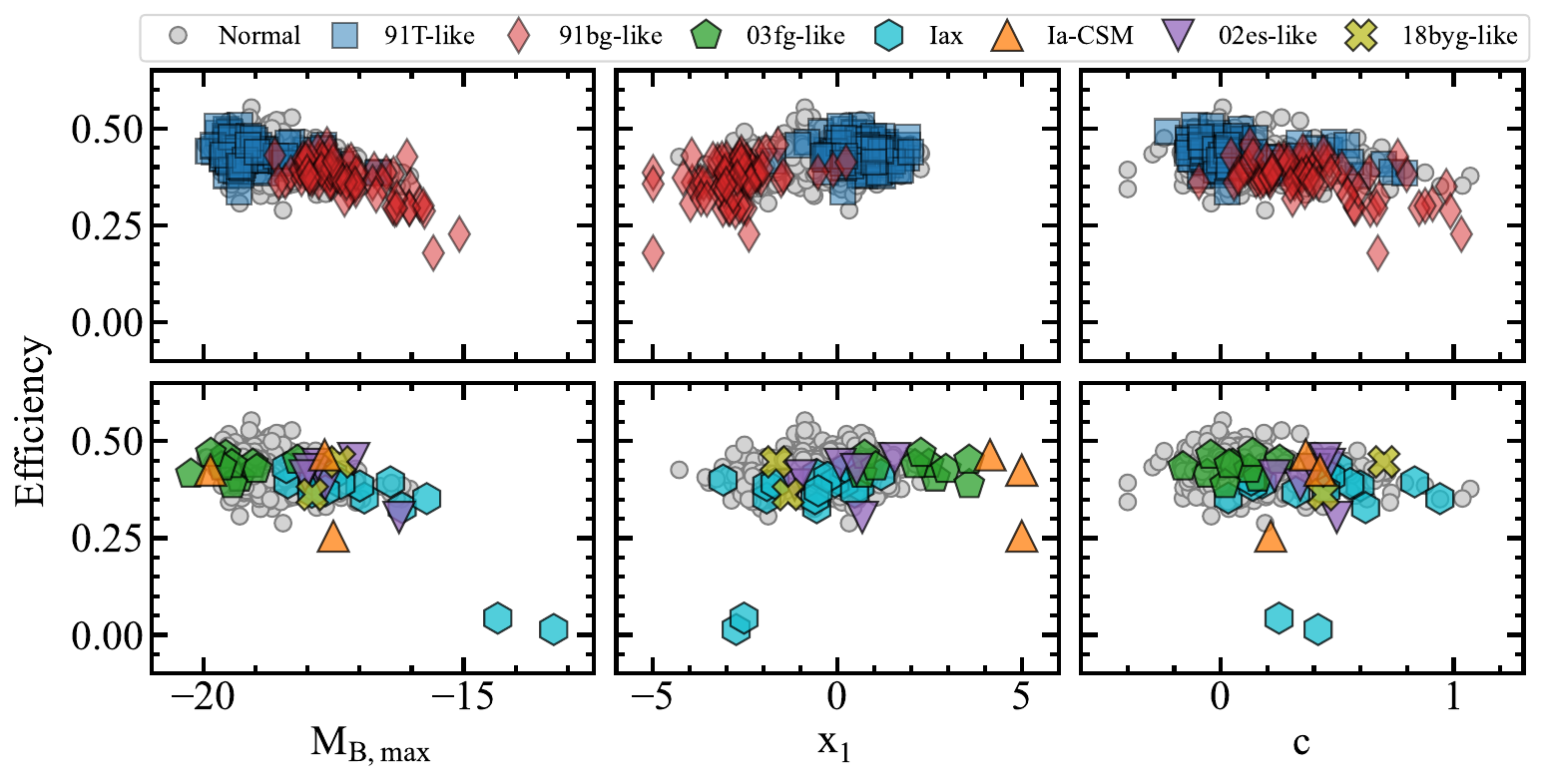}
    \caption{Efficiencies of the lccoverage\_flag=1 SNe Ia in our sample as a function of peak absolute magnitude in $B$ band (\textbf{left}), $x_{1}$ (\textbf{middle}) and $c$ (\textbf{right}). The different sub-types are presented in different symbols and colours, as shown in the legend.}
    \label{fig:efficiencies}
\end{figure*}

We now turn to the real events of ZTF SN Ia DR2. For each event with lccoverage\_flag=1, we have its fitted \texttt{SALT2} $x_{1}$, $c$, and apparent $B$-band peak magnitude (from $x_{0}$), and can calculate the absolute $B$-band peak magnitude from its redshift. We can then estimate its efficiency by linearly interpolating the simulation results in terms of $x_{1}$, $c$, and absolute $B$-band peak magnitude from above. We can now estimate the total number of events for each sub-class by correcting with the factor $1/eff$ to each event. We assume Poissonian errors for the number of events in each bin and for each sub-class. The efficiencies of our SNe Ia are presented in Fig.~\ref{fig:efficiencies}.

As it can be seen, the main factor affecting the efficiency is the intrinsic brightness of the transient: transients with absolute magnitudes less than $\sim-17$ have a constant efficiency of $\sim0.4$, while for fainter events the efficiency starts to drop. On the other hand, the width of the light curve and its colour (probed by $x_{1}$ and $c$, respectively) have a smaller effect and only at the extreme ends of the distributions. We note that a very small efficiency (thus a very large correction factor) is estimated for the faint SNe Iax in our sample. 

A shortcoming of our approach is the fact that the \texttt{SALT2} spectral template does not fully capture the light curve evolution of the non-normal events. Nevertheless, it generally describes (to the first order) the light curve evolution of most thermonuclear transients. Moreover, as we have shown above, for the ZTF observing strategy and for $z\leq0.06$, the main factors that contribute to the efficiency of detection for a transient is its intrinsic peak magnitude and its distance, because of the high and regular cadence alongside the deep magnitude limits. We do note though that, as it can be seen from Fig.~\ref{fig:x1_vs_dm15}, the decline rate in $g$-band for the fainter events, such as the 91bg-like, the 02es-like, and the Iax is faster than the normal SNe Ia for a given $x1$. This means that the efficiencies we calculate may be overestimated, thus the final rates for these events must be considered as a lower limit. Finally, in our \texttt{skysurvey} simulations we do not include any information on the host galaxy surface brightness and its contrast with the brightness of the transient. Based on the transient source detection efficiencies of PTF, the efficiency sharply drops for events with a host flux to transient flux ratio of $\gtrapprox0.6$ \citep[Fig. 6 from][]{Frohmaier2017ApJS}. This will primarily affect the intrinsically faint sub-classes and could result in an underestimate of the SN Iax rate, and to a lesser extent, the 91bg-like, 02es-like, and 18byg-like rates.

In the volume limited sample, for which we calculate our rates, 193 events ($\sim12.2\%$) belong to the not sub-typed group. Since we could not clearly identify their sub-type, we use our results from Sect.~\ref{sec:results_time_vs_color15} and particularly from Fig.~\ref{fig:color_rg_vs_t_rg}. As we have already mentioned, it appears that almost all of the not sub-typed events occupy the parameter space of normal/91T-like. We consider two cases; firstly, all not sub-typed events are normal SNe Ia, and secondly, a fraction of them are 91T-like and the remaining are normal. For the latter, we arbitrarily define the distinction at $(g-r)_{+15d}=0.3$, with events with bluer colour being 91T-like and with redder colour being normal. We adopt the mean of the above values as a conservative estimate.

\begin{figure}
    \centering
    \includegraphics[width=1\columnwidth]{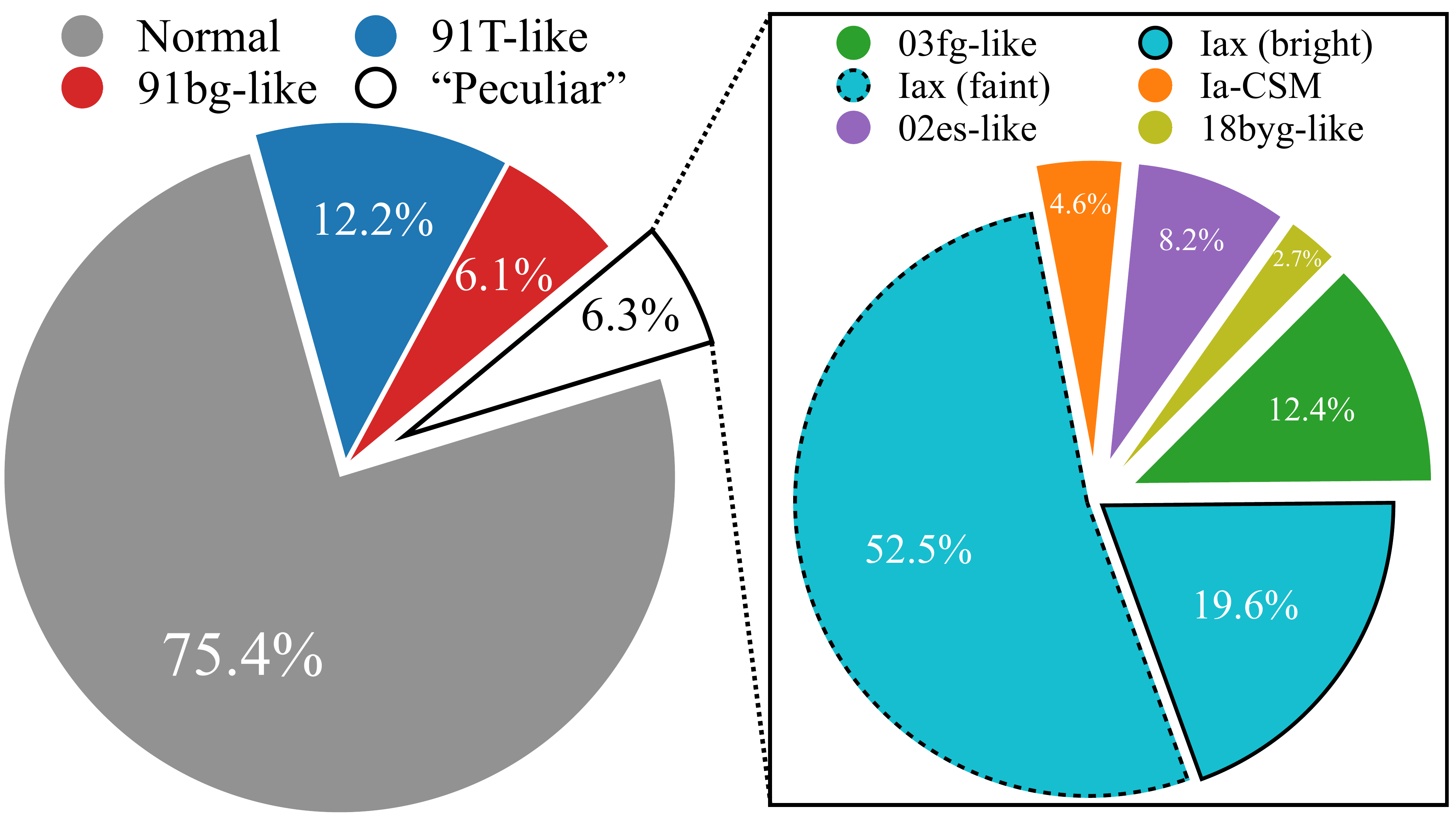}
    \caption{Pie charts of the efficiency-corrected observed fractions of the SN Ia sub-classes of the ZTF SN Ia DR2. The various sub-classes are presented with different colours, as shown in the legends.}
    \label{fig:fractions}
\end{figure}

Our efficiency-corrected observed fractions in our volume-limited sample indicate that the normal SNe Ia account for $75.4\pm3.5\%$, the 91T-like for $12.2\pm2.3\%$ (with $40.9\pm6.5\%$ of them being 99aa-like) and the 91bg-like for $6.1\pm0.8\%$ of SNe Ia. The remaining $6.3\%$ correspond to the rarer peculiar sub-classes, for which we estimate $0.8\pm0.2\%$ for the 03fg-like, $4.5\pm2.4\%$ for the Iax (split to $1.2\pm0.3\%$ and $3.3\pm2.5\%$ for the bright and faint ones, as described in Sect.~\ref{sec:results_phot_prop_iax}), $0.3\pm0.2\%$ for the Ia-CSM, $0.5\pm0.2\%$ for the 02es-like and $0.2\pm0.1\%$ for the 18byg-like SNe Ia. These final relative rates are shown as pie charts in Fig.\ref{fig:fractions}.

\section{Discussion}
\label{sec:discussion}

In this work, we performed a large-scale investigation of the ZTF SN Ia DR2 sample. We firstly described how we spectroscopically (sub-)classified 3,628 SNe Ia (Sect.~\ref{sec:classifications}) and how we modelled their light curves in a purely data-driven fashion (Sect.~\ref{sec:model_lc}). We then presented our analysis of their photometric (Sect.~\ref{sec:results_phot_prop} and \ref{sec:results_time_vs_color15}) and host-galaxy properties (Sect.~\ref{sec:results_hosts}), and we finally estimated their observed fractions (Sect.~\ref{sec:results_rel_rates}).

\subsection{The diversity of thermonuclear SNe Ia with ZTF}

While the photometric diversity of SNe Ia has been presented in previous studies, this work utilises the largest low-redshift transient survey, offering a homogeneous picture of the thermonuclear SN population. Figure 1 of \citet{Taubenberger2017hsn} presents a collection of SNe Ia in the brightness-evolution timescale phase space, revealing a clear separation of the peculiar population from the normal one (with the exception of the 91T-like). Our analysis shows similar results, but with some notable differences, stemming from the nature and origin of our sample. Firstly, the photometric band that was used in \citet{Taubenberger2017hsn} was the $B$-band, which is somewhat narrower and bluer (effective width and central wavelength of $\sim960$ and $\sim4,360\:\AA$) than our ZTF $g$ band ($\sim1,200$ and $\sim4,805\:\AA$, respectively), probing a different part of the SED. This is particularly important for the lower-luminosity sub-types (faint normal and 91bg-like events), where the absorption feature of \ion{Ti}{ii} is increasingly present at $4,300\:\AA$, allowing for the B-band to ideally 
probe that region. Moreover, the normal population presented there was the 185 SNe Ia from \citet{Hicken2009ApJ}, which were corrected for host extinction\footnote{\citet{Hicken2009ApJ} used MLCS \citep{Riess1996ApJ} to empirically estimate the total line-of-sight extinction.}, while the peculiar events were selected, well-observed, nearby SNe from the literature of various surveys, using numerous different telescopes and instruments. On the contrary, our sample originates from a single homogeneous survey, showing overlapping sub-type regions but outlining the real observed photometric diversity.

A method to identify non-cosmological SNe Ia from the photometry has been suggested by \citet{Ashall2020ApJ}, using the time difference between the maxima of two photometric bands (ideally a set of filters consisting of a blue and a red one, with significant separation in their effective wavelengths, such as the $g$ and $i$ bands) and the colour-stretch parameter \citep{Burns2014ApJ}. We propose a similar method that uses the time difference as in \citet{Ashall2020ApJ} and the colour of the two photometric bands 15 days after maximum brightness. Our results confirm the exclusively spectroscopic peculiarity of 91T-like events, as their photometric properties overlap with the brighter normal SNe Ia, with similar values of $t_{max}^{i-g}$ and colours. \citet{Ashall2020ApJ} finds similar $t_{max}^{i-g}$ for the normal SNe Ia, but somehow larger ($-1.12\pm0.47$) for the 91T-like, although their 91T-like sample is smaller. This potentially indicates a common explosion mechanism for the normal and the 91T-like, forming a continuum in the luminous region of SNe Ia (91T-like, 99aa-like and overluminous normal SNe Ia), with second-order progenitor scenario parameters \citep[e.g. the extent of $^{56}$Ni mixing into the ejecta,][]{Phillips2024ApJS} affecting the temperature and leading to the higher ionisation state we observe in the 91T-like spectra. Moreover, a similar continuum in the spectroscopic properties from bright to faint normal SNe Ia has been seen in \citet{Burgaz2024AA}. However, whether these trends can be attributed to a single explosion mechanism or binary (or both) with a varying parameter, or several progenitor scenarios are acting simultaneously and reproduce these trends, is still unclear.

Non-cosmological SNe Ia have generally redder $(g-i)_{+15d}$ colours, while the time difference $t_{max}^{i-g}$ indicates that the $i$ band peaks after the $g$ band (and in fact, the same thing for any band redward of $g$). This trend almost certainly originates from the source that powers their light curves. The 91bg-like SNe Ia show no distinctive $i$-band secondary maximum, with a faster decline in the $g$ band, due to the low-temperature explosion leading to a merging of the primary and secondary $i$-band maximum. The situation is more complicated for SNe Iax; under the partial-deflagration scenario \citep{Fink2014MNRAS}, widely thought to be the progenitor scenario of these events, the mixing of $^{56}$Ni in the outer layers sustains high-ionization states, and combined with the low expansion velocities observed, results in a longer diffusion timescale and the disappearance of the $i$-band secondary maximum. On the other hand, the $i$-band behaviour of the 03fg-like, 02es-like, and SNe Ia-CSM is driven by the interaction of the SN ejecta with CSM around the explosion site and the effective trapping of the radiation. The 03fg-like SNe Ia have strong indications of a carbon/oxygen-rich and hydrogen-poor CSM envelope that leads to longer diffusion timescales \citep{Dimitriadis2023MNRAS,Srivastav2023ApJ,Siebert2024ApJ}, while a connection between the 02es-like and the 03fg-like sub-classes has been demonstrated \citep{Srivastav2023ApJ,Hoogendam2024ApJ}, hinting for a common progenitor scenario, with variations in the mass of the CSM envelope and the mass of the synthesised $^{56}$Ni. Finally, the post-peak light curves of SNe Ia-CSM are dominated by the interaction of their ejecta with a hydrogen/helium-rich CSM, usually associated with a single degenerate scenario origin.

Focusing on the host galaxies of the peculiar SNe Ia, we find that specific sub-types tend to explode in particular environments. While these correlations have been explored in large samples of SNe Ia \citep[e.g.][]{Sullivan2010MNRAS, Kelly2010APJ, Galbany2014AA, Galbany2018ApJ}, very few studies have concentrated on the SNe Ia sub-classes \citep{Hakobyan2020MNRAS,Chakraborty2024ApJ} and particularly in a homogeneous and volume-limited sample. The host galaxy preference of the 91bg-like SNe Ia has been observed before \citep[see][and references therein]{Taubenberger2017hsn} and we confirm that they prefer higher mass and redder colours, with less star formation. On the other hand, the previously thought 91T-like preference for lower mass and bluer colours (thus higher star formation) is not statistically significant, similar to the findings of \citet{Phillips2024ApJS}. We generally find smaller masses and bluer colours for the 03fg-like and the SNe Iax, and similar host galaxies for the 02es-like and 18byg-like SNe Ia with the 91bg-like. The host galaxies of SNe Ia-CSM appear to be, intrinsically, the less diverse, with the notable exception of the single Helium-rich SN Ia-CSM. Nevertheless, future surveys with deeper limits will discover more peculiar SNe Ia, increasing the size of the sample and decreasing the statistical uncertainties.

\subsection{The intrinsic relative rate of thermonuclear SNe in a volume-limited sample}

We have estimated the relative rates of the SN Ia sub-classes of our volume-limited sample, after correcting for the efficiency of detection that affects primarily the fainter sub-classes. We estimate that $\sim75\%$ of the SN Ia explosions in the local universe are normal SNe Ia, with $\sim12\%$ and $\sim6\%$ corresponding to the traditional peculiar sub-types of 91T-like and 91bg-like, respectively. Comparing our results with the volume-limited sample of \citealt{Li2011MNRAS}, we find different fractions for the normal, 91T-like and 91bg-like SNe Ia ($70\%$, $9\%$ and $15\%$, respectively) but similar for SNe Iax ($5\%$), probably due to the targeted (to more massive galaxies) nature of the LOSS survey. We also find less normal and 91T-like SNe Ia than \citealt{Desai2024MNRAS} (89.5$\pm11.8$\% and 3.7$\pm0.8$\%, respectively) and identical fraction for the 91bg-like. We note though that in \citet{Desai2024MNRAS}, the normal SNe Ia contain events that they were unable to sub-classify (similar to our not sub-typed ones, for which we attempted a classification). When we perform the same assumption (i.e. include our not sub-typed events to the normal population), we find a normal population of $\sim78.7\pm$4.7\%, in agreement with \citet{Desai2024MNRAS} to $\sim0.85\sigma$, but with smaller uncertainty. On the other hand, the disagreement for the 91T-like population is significant, at the $\sim3.5\sigma$ level. These discrepancies can be due to lower-number statistics, the lack of certain sub-types in the study (Iax and 18byg-like are not considered, while it is not clear if 99aa-like events are included in the normal or the 91T-like sub-class), and/or limitations due to the survey used, as the ASAS-SN sample had single photometric band light curves and substantially shallower magnitude limit ($\sim17$ mag). The same explanation might be relevant for our increased percentages of the other rarer sub-types considered in \citet{Desai2024MNRAS}, as we find $0.8\%$ and $0.3\%$ compared to $0.1\%$ and $0.04\%$ for the 03fg-like and SNe Ia-CSM, respectively, although we find similar fractions for the 02es-like ($0.4\%$ to our $0.5\%$). We also note that our final SNe Iax rate relative to the normal SNe Ia (6.5$\pm$3.7\%) is generally consistent with the lower rate calculated by \citealt{Foley2013ApJ} (31$^{+17}_{-13}$\%) to the 1.8$\sigma$ level.

\section{Conclusions}
\label{sec:conclusion}

In this paper we presented an overview of the photometric diversity and relative rates of SNe Ia in the ZTF SN Ia DR2 (\citealt{Rigault2024aAA}, \citealt{smith2024AA}). Our main results are the following:

\begin{enumerate}
    \item The peculiar sub-classes of the ZTF SNe Ia sample follow the previously established trends in the brightness--light curve evolution phase space. The 03fg-like and SNe Ia CSM are generally brighter and slower; the 91bg-like and Iax are fainter and faster; and the 02es-like have normal light curve evolution timescales, but fainter. The small number of 18byg-like SNe Ia prevents us from drawing firm conclusions, but they generally appear to be fainter than the normal population.
    \item The 91T-like SNe Ia have photometric properties that are identical to those at the brighter end of the normal population; the differences are strictly spectroscopic.
    \item A method for identifying peculiar events in non-spectroscopic samples is measuring the differences of the timing of the maximum of a blue and a red photometric band and the colour at $+15$ days from peak brightness. All peculiar sub-types, apart from the 91T-like SNe Ia, are clearly separated in that phase space from the normal SNe Ia, and provide an additional tool for distinguishing cosmologically non-useful events.
    \item The properties of the host galaxies of the peculiar sub-classes follow the established trends seen in the literature. An exception is the 91T-like events, for which we do not find a statistically significant preference to occur in low mass and star-forming galaxies.
    \item The observed fraction of the cosmologically useful SNe Ia in the local Universe is $\sim87\%$, with $\sim75\%$ being normal and $\sim12\%$ 91T-like SNe Ia. The 91bg-like population corresponds to $\sim6\%$ of the total population. The remaining $\sim7\%$ of events belong to the rarer peculiar sub-classes of SNe Iax ($\sim4\%$, the most common peculiar sub-class), 03fg-like ($1\%$), Ia-CSM ($0.3\%$), 02es-like ($0.5\%$) and 18byg-like ($0.2\%$).
\end{enumerate}

While the peculiar thermonuclear explosions present a valuable opportunity to study exotic explosion scenarios and binary systems, their rare nature limits large-scale statistical analysis. The ZTF SN Ia DR2 provides the current best sample to perform these studies, and can provide the benchmark for the Vera Rubin Legacy Survey of Space and Time \citep[LSST;][]{Ivezic2019ApJ}. Nevertheless, there are still limitations. Standard spectroscopic classifiers, such as \texttt{SNID}, \texttt{SuperFit}, \texttt{GELATO}, and \texttt{DASH} \citep{Muthukrishna2019ApJ} are not sufficient to distinguish between certain sub-types, particularly for the high-redshift LSST objects, as few of them will get a spectrum and potentially have low spectral S/N. Human expertise is still essential to uncover subtle spectroscopic features and make educated guesses. While photometric classifiers \citep{Moller2020MNRAS,Gagliano2023ApJ} are expected to perform very well in identifying tens of thousands of cosmologically useful SNe Ia in the LSST era, the field of peculiar SN Ia studies may suffer from not being able to properly classify the outliers. Therefore, we encourage more spectral observations for nearby events, especially at early times when the spectroscopic peculiarities appear to be more prominent.

\begin{acknowledgements}

We thank the anonymous referee for helpful comments that improved the clarity and presentation of this paper. Based on observations obtained with the Samuel Oschin Telescope 48-inch and the 60-inch Telescope at the Palomar Observatory as part of the Zwicky Transient Facility project. ZTF is supported by the National Science Foundation under Grants No. AST-1440341 and AST-2034437 and a collaboration including current partners Caltech, IPAC, the Weizmann Institute of Science, the Oskar Klein Center at Stockholm University, the University of Maryland, Deutsches Elektronen-Synchrotron and Humboldt University, the TANGO Consortium of Taiwan, the University of Wisconsin at Milwaukee, Trinity College Dublin, Lawrence Livermore National Laboratories, IN2P3, University of Warwick, Ruhr University Bochum, Northwestern University and former partners the University of Washington, Los Alamos National Laboratories, and Lawrence Berkeley National Laboratories. Operations are conducted by COO, IPAC, and UW. SED Machine is based upon work supported by the National Science Foundation under Grant No. 1106171. The ZTF forced-photometry service was funded under the Heising-Simons Foundation grant \#12540303 (PI: Graham). This work was supported by the GROWTH project funded by the National Science Foundation under Grant No 1545949 \citep{Kasliwal2019pasp}. Fritz \citep{vanderWalt2019, Coughlin2020apjs} is used in this work. The Gordon and Betty Moore Foundation, through both the Data-Driven Investigator Program and a dedicated grant, provided critical funding for SkyPortal. GD, UB, MD, KM, and JHT are supported by the H2020 European Research Council grant no. 758638. This project has received funding from the European Research Council (ERC) under the European Union's Horizon 2020 research and innovation program (grant agreement n 759194 - USNAC). LG, AA and TEMB acknowledges financial support from the Spanish Ministerio de Ciencia e Innovaci{\'o}n (MCIN) and the Agencia Estatal de Investigaci{\'o}n (AEI) 10.13039/501100011033 under the PID2020-115253GA-I00 HOSTFLOWS project, and from Centro Superior de Investigaciones Cient{\'i}ficas (CSIC) under the PIE project 20215AT016 and the program Unidad de Excelencia Mar{\'i}a de Maeztu CEX2020-001058-M. LG acknowledges financial support from the Departament de Recerca i Universitats de la Generalitat de Catalunya through the 2021-SGR-01270 grant. KP acknowledges financial support from AGAUR, CSIC, MCIN and AEI 10.13039/501100011033 under projects PID2023-151307NB-I00, PIE 20215AT016, CEX2020-001058-M, and 2021-SGR-01270. TEMB acknowledges financial support from the European Union Next Generation EU/PRTR funds under the 2021 Juan de la Cierva program FJC2021-047124-I. Y-LK has received funding from the Science and Technology Facilities Council [grant number ST/V000713/1]. AAM is partially supported by LBNL Subcontract NO.\ 7707915. SD acknowledges support from a Kavli Fellowship and a Junior Research Fellowship at Lucy Cavendish College. This work has been supported by the research project grant “Understanding the Dynamic Universe” funded by the Knut and Alice Wallenberg Foundation under Dnr KAW 2018.0067 and the {\em Vetenskapsr\aa det}, the Swedish Research Council, project 2020-03444. LH is funded by the Irish Research Council under grant number GOIPG/2020/1387.

\end{acknowledgements}

%
%

\bibliographystyle{aa}
\bibliography{ztf_dr2_ia_pop_v3.bib}

\begin{appendix} 
\onecolumn

\section{Sample parameters and statistics}
\label{sec:sample_statistics}

\begin{table}
\caption{Summary of the cuts and the associated population numbers of our samples.}\label{tab:samples_numbers}
\centering
\begin{tabular}{l r|c c c c c c c c c } 
\hline \\[-0.5em]
Sample & & Normal & 91T-like & 91bg-like & 03fg-like & Iax & Ia-CSM & 02es-like & 18byg-like & not sub-typed \\
\hline \\
All & & \textbf{2511} & \textbf{292} & \textbf{92} & \textbf{29} & \textbf{23} & \textbf{14} & \textbf{8} & \textbf{4} & \textbf{655} \\

& Coverage cut & 1216 & 141 & 34 & -- & -- & -- & -- & -- & 234 \\
& Fit quality cut & 1143 & 133 & 32 & -- & -- & -- & -- & -- & 223 \\
& $c<0.3$ & 1093 & 126 & -- & -- & -- & -- & -- & -- & -- \\
gr & & \textbf{1093} & \textbf{126} & \textbf{32} & \textbf{24} & \textbf{17} & \textbf{6} & \textbf{6} & \textbf{2} & \textbf{223} \\
& Coverage cut & 197 & 24 & 10 & -- & -- & -- & -- & -- & 35 \\
& Fit quality cut & 189 & 23 & 10 & -- & -- & -- & -- & -- & 33 \\
gri & & \textbf{189}  & \textbf{23} & \textbf{10} & \textbf{7} & \textbf{6} & \textbf{1} & \textbf{3} & \textbf{1} & \textbf{33} \\
gr-zcut & $z\leq0.06$ & 404 & 45 & 30 & 8 & 15 & 1 & 5 & -- & 25 \\
gri-zcut & $z\leq0.06$ & 55 & 8 & 9 & 2 & 6 & -- & 2 & -- & 4 \\
\hline
\end{tabular}
\end{table}

\begin{table}
\caption{SN GP fit parameters.}\label{tab:samples_params}
\centering
\begin{tabular}{l |c c c c c c c}
\hline \\[-0.5em]
 & No K-corrections &  &  &  &  &  & ... \\
SN Name & MJD$^{peak}_{g}$ & MJD$^{peak}_{r}$ & MJD$^{peak}_{i}$ & g$^{peak}$ & r$^{peak}$ & i$^{peak}$ & ... \\
\hline \\
ZTF17aadlxmv & $58878.055\pm0.270$ & $58878.315\pm0.233$ & -- & $17.813\pm0.023$ & $17.862\pm0.024$ & -- & ...  \\
ZTF18aaaqexr & $58893.578\pm0.336$ & $58893.998\pm0.450$ & -- & $18.485\pm0.048$ & $18.486\pm0.048$ & -- & ...  \\
ZTF18aadzfso & $58887.663\pm0.324$ & $58888.693\pm0.209$ & -- & $18.180\pm0.028$ & $18.087\pm0.025$ & -- & ...  \\
ZTF18aagrtxs & $58212.640\pm0.150$ & $58212.810\pm0.290$ & -- & $16.197\pm0.009$ & $16.287\pm0.021$ & -- & ...  \\
ZTF18aaguhgb & $58215.504\pm0.121$ & $58216.949\pm0.183$ & -- & $18.931\pm0.023$ & $18.956\pm0.021$ & -- & ...  \\
\hline
\end{tabular}
\tablefoot{All parameters for all SNe in our sample are available at \href{http://ztfcosmo.in2p3.fr}{ztfcosmo.in2p3.fr} and at the CDS.}
\end{table}

\end{appendix}

\end{document}